%% file: cameraready.tex
\documentclass[acmlarge]{acmart}
\makeatletter
\newcommand{\confshort}{\acmConference@shortname}
\newcommand{\conffull}{\acmConference@name}
\newcommand{\confdate}{\acmConference@date}
\newcommand{\confloc}{\acmConference@venue}
\AtBeginDocument{
  \fancypagestyle{firstpagestyle}{
    \fancyhead{}%
    \fancyfoot[C]{}%
  }
  \fancyhf{}
  \fancyhead[LO]{\@headfootfont\shorttitle}%
  \fancyhead[RE]{\@headfootfont\@shortauthors}%
  \fancyhead[LE]{\@headfootfont\footnotesize \confshort, \confdate, \confloc}%
  \fancyhead[RO]{\@headfootfont\footnotesize \confshort, \confdate, \confloc}%
  \fancyfoot[C]{}%
}
\makeatother
\acmBooktitle{\conffull\@ (\confshort), \confdate, \confloc}

\usepackage[T1]{fontenc}
\usepackage{tipa}
\usepackage{dcolumn}
\usepackage{graphicx}
\usepackage{multirow}%
\usepackage{amsmath}%

\usepackage{amssymb}
\usepackage{amsthm}%
\usepackage{rotating}
\usepackage{makecell}
\usepackage{mathrsfs}%
\usepackage[table]{xcolor} 
\usepackage[title]{appendix}%
\usepackage{xcolor}%
\usepackage{textcomp}%
\usepackage{manyfoot}%
\usepackage{booktabs}%
\usepackage{algorithm}%
\usepackage{algorithmicx}%
\usepackage{algpseudocode}%
\usepackage{listings}%
\usepackage{float}
\usepackage{hyperref}

\usepackage{lscape}

\AtBeginDocument{%
  }

\copyrightyear{2026}
\acmYear{2026}
\setcopyright{cc}
\setcctype{by-nc-nd}
\acmConference[FAccT '26]{The 2026 ACM Conference on Fairness, Accountability, and Transparency}{June 25--28, 2026}{Montreal, QC, Canada}
\acmBooktitle{The 2026 ACM Conference on Fairness, Accountability, and Transparency (FAccT '26), June 25--28, 2026, Montreal, QC, Canada}
\acmDOI{10.1145/3805689.3812320}
\acmISBN{979-8-4007-2596-8/2026/06}

\begin{document}

\title[Addressing Auditing Pitfalls in ASR Technologies]{Addressing Auditing Pitfalls in Automatic Speech Recognition Technologies: A Case Study of People with Aphasia}

\author{Katelyn X. Mei}
\authornote{Both authors contributed equally to this research.}
\affiliation{%
 \institution{University of Washington}
 \city{Seattle}
 \state{Washington}
 \country{USA}}
\email{kmei@uw.edu}

\author{Anna Seo Gyeong Choi}
\authornotemark[1]
\affiliation{%
  \institution{Cornell University}
  \city{Ithaca}
  \state{New York}
  \country{USA}}
\email{sc2359@cornell.edu}

\author{Hilke Schellmann}
\affiliation{%
  \institution{New York University}
  \city{New York}
  \state{New York}
  \country{USA}}
\email{hilke.schellmann@nyu.edu}

\author{Mona Sloane}
\affiliation{%
  \institution{University of Virginia}
  \city{Charlottesville}
  \state{Virginia}
  \country{USA}}
\email{mona.sloane@virginia.edu}

\author{Allison Koenecke}
\affiliation{%
  \institution{Cornell Tech}
  \city{New York}
  \state{New York}
  \country{USA}}
\email{koenecke@cornell.edu}

\renewcommand{\shortauthors}{Mei \& Choi et al.}

\begin{abstract}
Automatic Speech Recognition (ASR) systems' growing use warrants robust auditing approaches to ensure equitable transcription quality, especially for people with speech disorders like aphasia who disproportionately depend on ASR. While academic and industry audits have revealed performance disparities across user populations, standard auditing practices often overlook nuances that risk masking harm to marginalized groups. We identify three common pitfalls in standard ASR audits: (1) adhering to one method of text standardization, which can mask variance in ASR performance and ignore the standardization preferences of marginalized communities; (2) displaying high-level demographic findings without considering performance disparities by nuanced intersectional subgroups, or conditioning on relevant acoustic properties; and (3) reporting only one gold-standard metric (Word Error Rate), which inadequately quantifies common generative AI errors like hallucinations. We propose a holistic auditing framework addressing these pitfalls, and in a case study of six popular ASR systems, find consistently worse ASR performance for speakers with aphasia relative to a control group. We call on practitioners to implement these robust, community-driven ASR auditing practices better suited for the rapidly changing ASR landscape.

\end{abstract}



\maketitle

\section{Introduction}
Automatic Speech Recognition (ASR) technologies transcribe spoken audio into written text and are increasingly integrated into daily life, from powering virtual assistants to generating meeting transcriptions. In high-stakes settings like healthcare, job interviews, and court reporting, ASR inaccuracies can cause direct harm, particularly to marginalized speakers including African American English speakers \cite{koenecke2020racial}, people who stutter \cite{lea2023user}, and refugees \cite{bircan2024machine}. Regular examination of state-of-the-art ASR services is critical to investigate ASR-based biases across user populations~\cite{koenecke2025techbrief}. Academic audits have revealed accessibility and performance disparities for marginalized populations by race \cite{koenecke2020racial, martin2023bias, mengesha2021don}, gender \cite{feng2021quantifying, feng2024towards, attanasio2024multilingual}, and medical condition \cite{moore2018whistle, lea2023user, HidalgoLopez2023, Zhao2024}, while ASR providers increasingly self-audit, reporting their performance relative to competitors~\cite{assemblyai2024,hughes2023ursa,radford2023robust,deepgram2024}.

Standard audits compare a ``gold standard'' ground truth audio transcription to ASR output \cite{Sloane2023, sloane2026case}, but often overlook nuances related to specific user communities, risking results that lack robustness or fail to account for marginalized groups' experiences \cite{barocas2021designing}. Most ASR audits follow a standard procedure~\cite{radford2023robust,lea2023user,koenecke2020racial}: text is ``cleaned'' in a standardized way (removing capitalization, punctuation, filler words, etc.) and compared to a similarly-cleaned ground truth; performance is quantified by a single error rate metric; and high-level comparisons may be made between the potentially harmed populations and a control group. However, we argue that each of these steps must be updated to improve auditing robustness, especially given advances in generative AI and the possibility of cherrypicked results in industry audits~\cite{gmfusAuditWashingAccountability}.

\begin{figure}[!t]
    \centering
    \includegraphics[width=\linewidth]{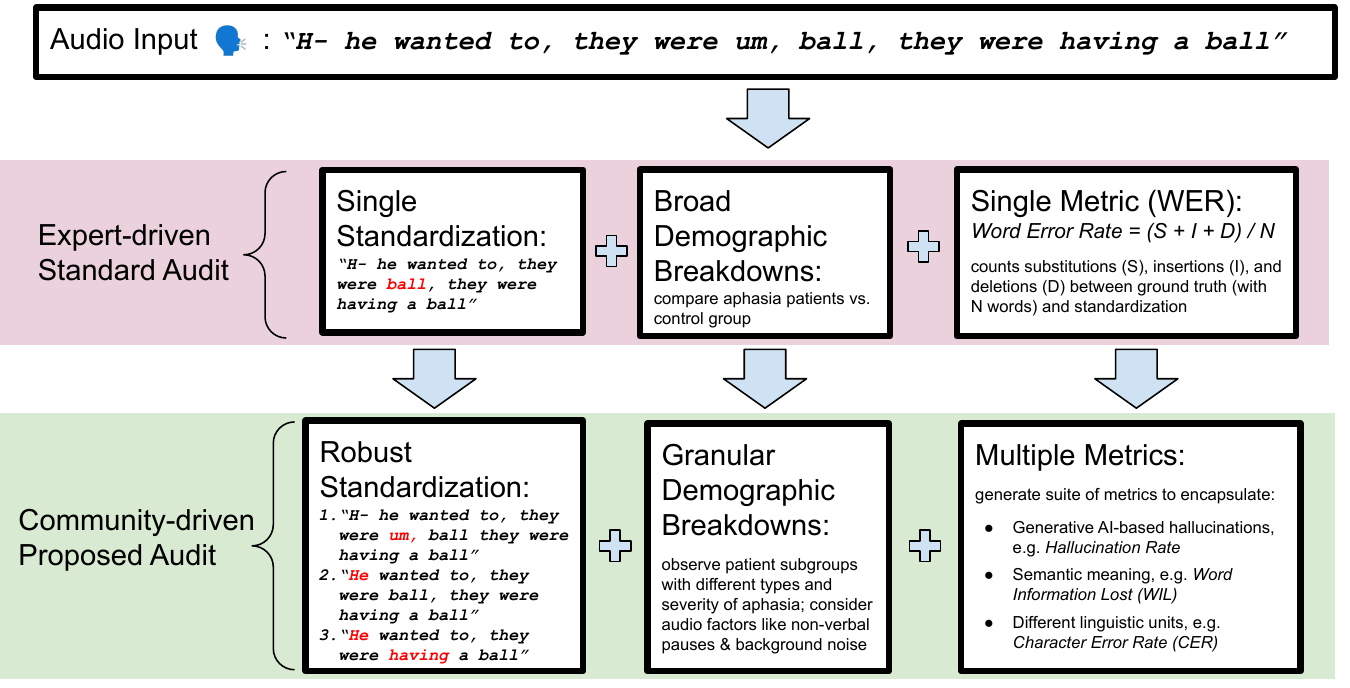}
    \caption{A standard audit takes an audio input, uses a single text standardization method (e.g., removing filler words like ``um''), examines only high-level group differences in performance (e.g., between only treatment and control groups), and quantifies performance using a single metric (e.g., WER). In contrast, we propose an auditing framework that is community driven. First, we consider multiple versions of text standardizations as a robustness check, taking into account the preferences of the relevant demographic communities (such as removing stutters). Second, we consider more granular and/or intersectional demographic subgroups as points of comparison. Third, we consider a suite of metrics to measure ASR performance.
    }
    \label{fig:audit_pipeline}
\end{figure}

We identify three core pitfalls pertaining to auditing ASR that warrant attention from the broader research community: (1) absence of robust text standardization experiments rarely considering community preference; (2) lack of consideration of nuanced demographic subgroup and audio characteristics; and (3) overemphasis on single evaluation metrics. We address each issue with our proposed recommendations for community-based auditing procedures (\autoref{fig:audit_pipeline}): first, different text standardization methods should be explored, and performance variance across methods be reported, since communities may prefer different transcription approaches. Second, studies should examine demographic group heterogeneity, considering intersectional features and covariates like background noise \cite{casey2017race, dale2015socioeconomic} and pauses \cite{geng2022speaker, romana2024fluencybank}.
Third, auditors should report multiple metrics rather than relying solely on the \emph{Word Error Rate (WER)}, which fails to capture semantic meaning or nuanced errors like hallucinated transcriptions \cite{koenecke2024careless}. 

Our study focuses on ASR performance for American English spoken by individuals with aphasia, a language disorder caused by brain damage affecting communication \cite{benson1996aphasia}. Aphasia affects 0.06\% of the population \cite{Code2011}, significantly impacting lives through work and social challenges \cite{graham2011aphasia, fama2022subjective}. We center aphasia patients because they are intersectionally disadvantaged: typically over 50 years of age \cite{nguy2022representation}, underrepresented in speech datasets, and disproportionately reliant on ASR technology \cite{wallacewells2024fetterman}.
Aphasia frequently impairs writing and typing due to its effects on language production more broadly \cite{lee2024typing}, making voice-based interfaces and ASR-powered tools an alternative accessible modality for communication and documentation \cite{nunez2023use}. Conversely, failures in ASR transcription carry amplified consequences for aphasia patients, who may have fewer alternative (e.g., text-based) modalities to fall back on.
We further highlight that ASR-generated \emph{hallucinations}---a kind of transcription error consisting of entirely fabricated phrases inserted into ASR output \cite{koenecke2024careless}---are particularly consequential for aphasia patients, who may frequently need to use ASR in high-stakes contexts (such as medical AI scribes). Hallucinated content in these downstream transcriptions, such as in medical documentation, can directly mislead doctors and caregivers, or misrepresent patients' intended speech \cite{burke2024}.

In our case study on aphasia speakers, we use AphasiaBank data \cite{aphasiabank} containing spontaneous speech interviews with aphasia and control group speakers, survey the aphasia community about ASR experiences and preferences, and test six popular services: Amazon, AssemblyAI, Google, Microsoft, OpenAI, and Rev AI. 
Our contributions include: (1) using standard auditing procedures from prior ASR fairness research \cite{koenecke2020racial, Zhao2024} to compare performance across six ASR services between speakers with and without aphasia, finding significantly worse WERs for speakers with aphasia (\S\ref{sec2.1}); (2) engaging with the aphasia community through surveys and interviews to better understand their lived experiences and ASR preferences (\S\ref{sec:community}); and (3) identifying three critical pitfalls in standard auditing practices, with empirical demonstrations of how addressing each pitfall yields more robust audit findings (\S\ref{sec:pitfall-standardization}---\S\ref{sec:pitfall-relying_on_wer}). We find that, while the standard audit reveals ASR disparities, it treats people with aphasia as monolithic and relies solely on WER, potentially masking nuanced performance disparities that---when uncovered---can inform technical improvements to ASR systems more broadly.

\section{Methods}

\subsection{Data Sources and Sample}\label{sec:aphasiabackground}

For our empirical case study, we used data from AphasiaBank \cite{macwhinney2011aphasiabank}, a publicly available repository containing audio recordings and transcriptions of structured interviews with people with aphasia and control participants, where participants perform narrative tasks such as story retelling and picture description. 
We analyzed a total of 551 Aphasia interviews and 347 Control interviews, downloaded directly from the AphasiaBank website. Participant demographic information provided by AphasiaBank includes aphasia or control group membership, age, gender, race/ethnicity, years of education, employment status, primary language, vision and hearing condition, and clinical impression of aphasia type if they had aphasia. Specific aphasia types were diagnosed by the clinicians conducting the interviews based on the Boston Diagnostic Aphasia Examination, and their speech-related characteristics are represented in Appendix~\ref{sec:aphasia-category}; we also obtained participants' Western Aphasia Battery (WAB) scores---a clinical measurement of language functioning.\footnote{Since WAB score ranges from 0 to 100, with a lower WAB score signifying a more severe status of aphasia, control participants were given a WAB score of 100 for statistical analysis.} We restricted our participant sample to individuals aged 18 years or older and for whom English was their first language.
All interviews in the present study were conducted in English; the AphasiaBank corpus contains speakers of various English dialects, though specific dialect information is not systematically documented in participant metadata.
Participant demographics are reported in Appendix Table~\ref{tab:overall_participant_demographics}.

Based on the raw transcription of the interviews provided by AphasiaBank, which transcribed the data following the CHAT transcription convention \cite{macwhinney2021tools}, we segmented each participant's speech into phrases uttered in one turn -- i.e., everything spoken by the interviewee between intermittent questions and responses from the interviewer. Interviews with more than one interviewee were discarded from the analysis. This resulted in 4,342 aphasia  audio segments and 1,843 control audio segments from 321 aphasia and 216 control speakers, respectively. Speakers with aphasia had an average segment duration of 38.8 seconds, speaking on average 27 words (after removing fillers); control speakers had an average segment duration of 71.1 seconds, speaking on average 100 words.

We refer to our ``ground truth transcriptions'' as those based on the AphasiaBank-provided transcriptions, which we programmatically cleaned to remove unintelligible words words and create standardized reference texts for WER calculations. Detailed procedures for data filtering, audio segmentation, and ground truth cleaning appear in Appendix~\ref{sec:cleaning-details}.

\subsubsection{Propensity Score Matching}\label{method:matching}

To enhance the robustness of our statistical analysis, we conducted dataset matching procedures with the R (Version 4.4.1) \textit{MatchIt} package. To obtain balanced samples between aphasia and control groups, we implemented propensity score matching using nearest neighbor matching with a caliper score of 0.13 to ensure a comparable distribution of age, gender, race/ethnicity, years of education, mean background noise level, and word count of ground truth for audio segments (described further in Appendix \S \ref{sec:acoustic-features}). This matching process resulted in 1,224 audio segments for the aphasia group and 1,224 audio segments for the control group.

We report results on both the full dataset and matched subsample throughout; findings are consistent across both samples. Participant demographic comparisons on matched and unmatched samples appear in Appendix Table~\ref{tab:pre-post-matching-count}, and standardized mean differences for covariates before and after propensity matching appear in Appendix Figure~\ref{fig:loveplot}.

\subsection{ASR System Selection}

We evaluated six widely-used commercial ASR services: Amazon AWS, AssemblyAI, Google Cloud Speech-to-Text, Microsoft Azure, OpenAI Whisper, and Rev AI. Amazon, Google and Microsoft were selected for being leading major tech companies and have been widely used in prior audits \cite{koenecke2020racial, Zhao2024, HidalgoLopez2023}, while AssemblyAI, OpenAI, and Rev AI represent prominent third-party services frequently utilized for specialized ASR applications, also employed in previous audits \cite{kuhn2024measuring, attanasio2024multilingual, heuser2024quantification}. 
For each ASR service, data collection (i.e., generation of automated transcriptions) for the aphasia group and the control group was done on the same day. 
We used default model configurations for each service with two exceptions: For Microsoft Azure, we specifically used the Azure Continuous model, as initial testing revealed the default model prematurely terminated transcription during long pauses characteristic of aphasia speech. For Google, we used the Chirp model as it represents their state-of-the-art general-purpose ASR system at the time of our experiments\footnote{All automated transcripts were collected during the first two weeks of October 2023. Google has since released Chirp 2 and Chirp 3, with the original models now classified as legacy systems.}.
All ASRs were confirmed to opt-out of data collecting services to ensure that our speech-to-text experiment data would not be used for training by the ASR models. 
Implementation details including API configurations and error handling procedures appear in Appendix~\ref{sec:asr-details}.

\subsection{Word Error Rate (WER)}\label{sec:wer_def}

Our primary evaluations of ASR performance use the industry standard WER metric, calculated as $\text{WER} = \frac{S + I + D}{N}$ where items in the numerator refer to the number of changes between the ground truth and ASR-generated transcriptions ($S$ = substitutions, $I$ = insertions, $D$ = deletions), and $N$ counts the total number of words in the ground truth transcription. A WER of 0 implies perfect transcription; lower WER is better. WER is the de facto standard in ASR evaluation, enabling comparison with prior audits. We additionally calculate WER at the group level (e.g., across speakers with aphasia) by aggregating all transcriptions within each group as if they were a single audio file---effectively a weighted WER---following \citet{ji2023survey}. We also report unweighted WER (averaging WER across individual audio files) in the Appendix; findings are consistent across both aggregation methods. We introduce non-WER ASR evaluation metrics in \S\ref{sec:pitfall-relying_on_wer}.

\subsection{Community Engagement}

To center the preferences and experiences of people with aphasia in our audit methodology, we conducted community engagement through surveys and interviews. Boston University Conversation, Health, Art, Technology (C.H.A.T)\footnote{\url{https://www.bu.edu/cbr/chat/}} is a resource group run by the Boston University Center for Brain Recovery. C.H.A.T. runs weekly sessions for the group members -- individuals with aphasia -- and provides a welcoming and supportive community via resource dissemination and group conversations about relevant research, therapies, or technology.
To ensure ethical and accessible engagement with participants with aphasia, the research team completed training in Supported Conversation for Adults with Aphasia (SCA\texttrademark), a standardized methodology developed by The Aphasia Institute for facilitating communication with and obtaining meaningful consent from individuals with aphasia \cite{kagan1998supported}.
We recruited seven C.H.A.T. participants to answer an online survey conducted through Qualtrics asking open-ended questions about their interactions with and opinions about ASR systems, which are discussed in \S \ref{sec:community}. 

\section{Standard Auditing Procedure}\label{sec2.1}

Following standard audit procedures, we examined WER across the six ASR services on our AphasiaBank dataset. 
For all six ASR systems, we conducted Mann-Whitney U tests to measure WER differences between aphasia and control speakers and applied the Benjamin-Hochberg method for multiple-testing correction. 
ASR transcriptions of aphasia speakers consistently yielded worse WERs (6-10 percentage points higher) than control speakers across all services  (see Appendix \autoref{tab:wer_weighted_unweighted} and \autoref{wer_unmatched_weighted}), revealing statistically significant (p-value $<0.001$) disparities. WERs for Microsoft were worst (aphasia---0.17, control---0.09) while RevAI performed best (aphasia---0.12, control---0.06). Results are comparable when comparing unweighted WER averages (see \S\ref{tab:wer_weighted_unweighted}).
Unweighted WER averages are consistently 4-9 percentage points higher than the weighted WER averages; regardless of weighting aggregation, the difference in WER between control and aphasia groups is consistent (a difference of 6-12 percentage points in both cases). Findings were also consistent when using 2,448 propensity-matched audio files (see \S\ref{method:matching}), balancing age, gender, education, and race/ethnicity (see Appendix \autoref{tab:wer_test_results}).

While these results highlight ASR underperformance for aphasia speakers, this standard analysis treats them as monolithic, ignoring variability in speech types and transcription preferences while focusing solely on WER. We address three key pitfalls in the following subsections by: (1) challenging assumptions in ground-truthing through different text standardization approaches; (2) revealing performance disparities across aphasia types and demographics; and (3) expanding beyond WER to explore additional evaluation metrics for ASR performance.

\section{Pitfall 1: Text Standardization Method}\label{sec:pitfall-standardization}

The first pitfall in current auditing studies occurs in the often-overlooked data pre-processing stage, where both ASR providers (through default output settings) and auditors (through subsequent cleaning) make standardization choices: when `cleaning' the ASR-generated transcriptions and ground truth text before calculating the WER. There is precedent for ASR providers adopting the cleaning protocol norms from academia: for example, the open-source package for text standardization used by OpenAI's Whisper~\cite{radford2023robust} was based on cleaning steps from an earlier academic study \cite{koenecke2020racial}. Across prior ASR audit studies, the concept of data cleaning is rarely, if at all, defined; it nearly always involves some form of standardization that lowercases all text and removes punctuation \cite{Ji2023}, but can also involve standardizing contractions (e.g., ensuring that `gonna' and `going to' are considered as equivalent transcriptions), removing filler words (such as `um' or `uh'), etc.

However, the default choices in standardization vary across different ASR services, meaning special attention must be paid to ensuring outputs across services are comparable. In \autoref{tab:groundtruth-step}, we provide an example focusing on one sample sentence uttered by an aphasia speaker in our corpus. When transcribing this specific sentence, we exemplify five different methods of standardization, each of which is valid and defensible. Existing industry distinctions between ``clean'' and ``verbatim'' transcripts are broad~\cite{amberscript_transcription_guidelines}, e.g., respectively,  as either ``grammatically correct'' or focused more on notation for external audio or paralinguistic features complying with consumer demands. These correspond to whether one would transcribe fillers, fragments, repeated words, or repeated phrases---or none of the above. Depending on which standardization method is chosen and applied to both the ground truth and ASR-generated transcription, the ensuing WER could change (since the number of words in the ground truth---i.e., the denominator of WER---could be different), thereby potentially changing audit results. The default settings of an ASR service may dictate the transcription cleaning method used by the majority of downstream users. For example, at the time of this study, while most ASR services automatically removed filler words from transcriptions, Amazon AWS did not. AWS also did not remove word fragments (when only the beginning of a word---but not the full word---is uttered), unlike most other ASR services. While most ASR services transcribed verbally repeated words, OpenAI Whisper might automatically remove these repetitions. While some services offered parameters for text standardization within the API, many did not. Some newer variants of ASR services even focus specifically on improving ``verbatim transcriptions,'' which ensure that filler words as originally stated are correctly transcribed~\cite{wagner2024crisperwhisper} rather than potentially arbitrarily removed.
We raise two interrelated standardization concerns in audit studies: ASR services don't consult speech communities' transcription preferences, yielding user-misaligned standardizations; and, studies rarely vary standardization methods, hindering robustness assessments across user preferences.

\begin{table}[htbp]
\caption{Examples of text standardization variants (iteratively applied each row) and whether the ASR services used during this study were observed as performing each standardization (denoted with checkmarks). Parentheses denote standardization only sometimes output by the ASR (but not as a default setting).}
\label{tab:groundtruth-step}
\centering
\footnotesize 
\begin{tabular}{p{5cm} p{5.8cm} *{6}{c}}
\toprule
\textbf{Cleaning Step} & \textbf{Example Sentence} & 
\rotatebox{90}{\parbox{2cm}{\centering Amazon AWS}} & 
\rotatebox{90}{\parbox{2cm}{\centering AssemblyAI}} & 
\rotatebox{90}{\parbox{2cm}{\centering Microsoft Azure}} & 
\rotatebox{90}{\parbox{2cm}{\centering Google Chirp}} & 
\rotatebox{90}{\parbox{2cm}{\centering OpenAI Whisper}} & 
\rotatebox{90}{\parbox{2cm}{\centering Rev AI}} \\
\midrule
Original & H- h- he he wanted to, they were um, ball they were having a ball. &  & & & & & \\[0.5ex]
Removing fillers (RF) \cite{kaushik2010automatic} & H- h- he wanted to, they were ball, they were having a ball. & & \checkmark & (\checkmark) & & \checkmark & \\[0.5ex]
Removing fragments (RFF) \cite{mimura2021end} & He he wanted to, they were ball, they were having a ball. & (\checkmark) & \checkmark & \checkmark & \checkmark & \checkmark & \checkmark \\[0.5ex]
Removing repeated single words (RFFR) \cite{sheikh2022machine} & He wanted to, they were ball, they were having a ball. & & \checkmark & \checkmark & & \checkmark & \\[0.5ex]
Removing repeated phrases (RFFRR) \cite{sheikh2022machine} & He wanted to, they were having a ball. & & (\checkmark) & (\checkmark) & & (\checkmark) & \\
\end{tabular}
\end{table}

\subsection{Incorporating Community-based Preferences}\label{sec:community}

Technology design researchers advocate for stakeholder inclusion in decision-making processes via ``Participatory AI''~\cite{kensing1998participatory,lee2020human,delgado2023participatory}, yet ASR audit practices often overlook user preferences despite directly impacting diverse users. For aphasia speakers saying \emph{``H- h- he he wanted to, they were um,''} text-cleaning choices range from verbatim transcription, to standardized meaning-preserving transcriptions like \emph{``He wanted to, they were.''} 
While existing ASR standardization defaults aren't necessarily unfaithful to the original audio file, decisions about how transcriptions are standardized should be made with different stakeholder preferences in mind.

Situated in aphasia speakers' everyday usage of ASR, we consider their preferences for output standardization as guidance for our auditing analysis, stipulating that their preferred ASR transcription style ought to constitute the ground truth. This community-first approach in AI auditing is particularly relevant for ASR used by people with disabilities, following the ``nothing about us without us'' principle \cite{charlton1998nothing, costanza2023nothing}. We performed in-depth qualitative interviews and quantitative online surveys on recruited participants with aphasia. We recruited seven C.H.A.T. participants to answer an online survey conducted through Qualtrics asking open-ended questions about their current view of ASR systems, of whom six additionally answered questions regarding their transcription preferences among the options listed in \autoref{tab:groundtruth-step}. Each of the seven participants answered how frequently they used ASR systems: rarely (1 participant), sometimes (1 participant), often (2 participants), or frequently (3 participants). They additionally answered how well ASR services performed for them; surprisingly, we found that only one participant answered ``not well at all''. The vast majority---5 participants---answered ``moderately well'', and one participant answered ``very well.'' Of the participants who answered questions regarding transcription standardization preferences, common themes included the desire for having an option for users to choose among transcription types, and the notion of different preferences depending on progression of aphasia (e.g., less standardization being necessary when a patient has had more years of speech therapy). We are conducting a parallel interview study on a larger set of aphasia participants for more in-depth qualitative discussion of their lived experience with aphasia and their ideals for inclusive ASR technology.
Our sample size of seven participants, while small, follows the practices of recent papers similarly exploring fairness in technology with only 10-13 interviewees~\cite{deng2022exploring, gu2021understanding, deng2023understanding,caine2016local}, and is further restricted given the low incidence rate of aphasia (estimated at 0.06\%~\cite{Code2011}).

All six participants preferred the most extensively-cleaned text standardization option provided (i.e., \emph{``He wanted to, they were having a ball''} in \autoref{tab:groundtruth-step}---removing filler words, fragments, and word and phrase repetitions). Qualitatively, these participants were in strong agreement that the most condensed version of the transcription was that which most accurately reflected their intended speech. We note that standardization preferences could vary by purpose and user group of ASR application. Though the dominant view was a preference for the cleanest standardization option, our community-based discussions revealed a willingness to consider that less ``cleaned'' standardizations could be used for speech therapy -- in particular, for tracking speech improvement over the course of aphasia patients’ recovery journeys. And, while aphasia patients using ASR services for everyday tasks may prefer the most-condensed standardization approach, medical doctors diagnosing aphasia or speech pathologists tracking rehabilitation progress may instead prefer verbatim transcriptions that include repeated words or phrases (crucial for monitoring medical conditions \cite{le2018automatic, fraser2013automatic, fraser2014automated, stemberger2020phonetic}).

\subsection{Reporting Performance Variance Across Standardization Methods}

\begin{figure}[!h]
    \centering
    \includegraphics[width=0.8\linewidth]{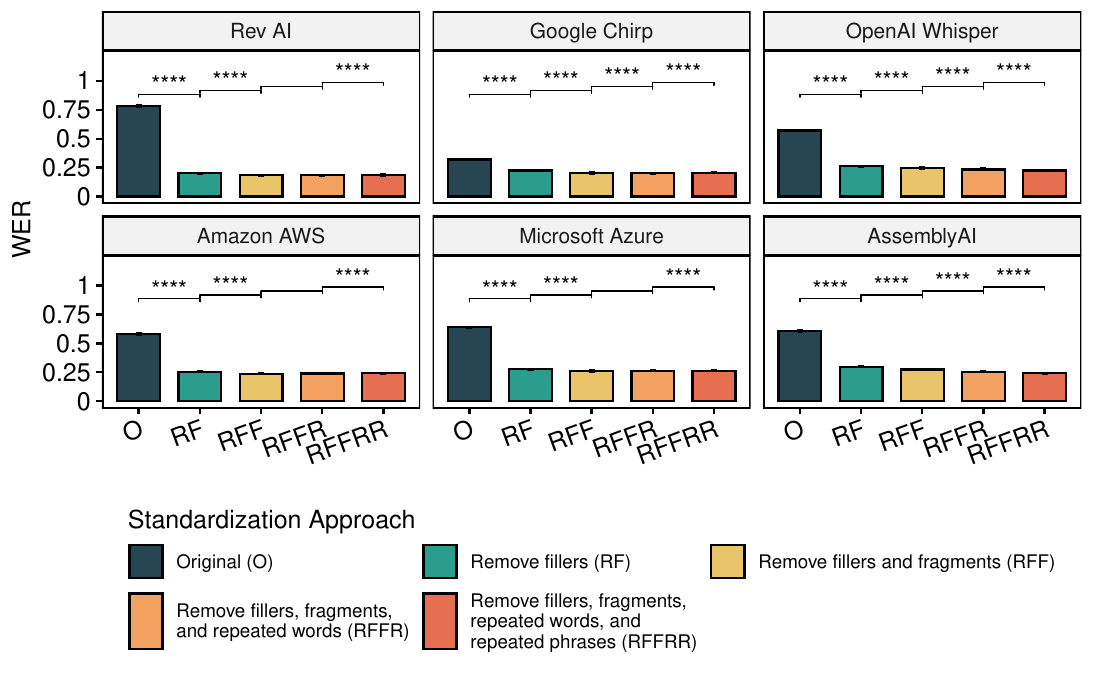}
    \caption{On a total sample of 4,342 aphasia files, we find that different standardization approaches (as defined in \autoref{tab:groundtruth-step}) yield different WER results for aphasia speakers. Removing fragments in addition to filler words significantly reduces WERs across all ASRs. However, half of ASR services (excepting OpenAI Whisper, Google Chirp, and AssemblyAI) did not yield statistically significant differences in WER when additionally removing repeated words, a method that more closely aligns with aphasia speaker preferences for text standardization. (Significance symbols for Wilcoxon signed rank tests: $ *: p<0.05; **: p <= 0.01;***: p <= 0.001 ;****: p <= 0.0001$.) See Appendix Tables \ref{tab:wer_standardization} and \ref{tab:mapsswe_aphasia} for further numerical tabulations.}
    \label{fig:groundtruth_comparison}
\end{figure}

Based on the surveyed aphasia speakers' responses aligning with most extensively-cleaned text standardization, our previously-presented ``standard'' audit results adhered to this standardization method (removing filler words, fragments, and phrase repetitions). 
However, we further propose that audits additionally showcase the variance in results across different levels of text standardization. For both aphasia and control audio segments across all six ASR systems, Kruskal-Wallis tests with multiple testing correction (Holm-Bonferroni) show statistically significant differences in WER across different standardization approaches, as shown in Appendix Table~\ref{tab:kw_test}  ($p<0.001$). Further pairwise comparisons---via Wilcoxon signed rank tests---among different levels of text standardization are shown in \autoref{fig:groundtruth_comparison}, indicating that WER for aphasia speakers improves significantly when removing filler words, and again when removing fragments. 

Choice of standardization method can alter both the reported WER, and the rank ordering of ASR services in performance audits. To assess if rank ordering of ASR differs across different levels of standardization, we adopted the Matched-Pair Sentence Segment Word Error (MAPSSWE) tests~\cite{gillick1989some}, a standard statistical test designed to assess differences in WER between systems for the same segments. The MAPSSWE tests on aphasia speakers revealed that standardization choices can reverse performance rankings (see Appendix Table~\ref{tab:mapsswe_aphasia}): for example, OpenAI Whisper significantly outperforms Amazon AWS with minimal standardization ($p < 0.001$), but Amazon AWS significantly outperforms Whisper with extensive standardization ($p = 0.011$).  
Robustness checks for the control group are reported in Appendix \autoref{fig:groundtruth_comparison_control}.

Strikingly, in half of the ASR services tested, there was no significant difference in average WER for aphasia speakers between removing fillers and fragments---RFF, and additionally removing repeated words---RFFR (as illustrated in \autoref{fig:groundtruth_comparison}, wherein pair-wise comparisons between different standardization approaches used Wilcoxon rank-sum tests). In these cases, aligning with aphasia speaker preferences for text standardization (removing word-level stutters) does not yield a significantly different WER. In all standardization approaches among aphasia patients, average WERs are within 2 percentage points of each other (findings are comparable among control patients, see Appendix \autoref{tab:wer_standardization}). This exemplifies that the choice to apply a community-based standardization---legitimized by ASR users, as opposed to externally-imposed by the ASR service themselves---is not necessarily at odds with attaining a competitive WER. 

We note that the community-preferred standardization method (RFFRR---i.e., removing fillers, fragments, repeated words, and repeated phrases) serves as the methodological foundation for all WER calculations reported in the ensuing sections (Pitfalls 2 and 3). That is, the subsequent findings are predicated on a preprocessing choice that was legitimized by the aphasia community rather than imposed by external convention---a distinction that we consider central to the community-based auditing framework we propose.

\section{Pitfall 2: Analysis of Overly-Broad User Groups} \label{sec:pitfall-demographic}
The second pitfall in ASR auditing studies is the tendency to perform only aggregate-level comparisons (e.g., comparing a treatment group to a control group). These analyses---often performed by broad demographic group--- ignore two important considerations discussed in the following subsections: that treatment groups are not monolithic, and that other important covariates may play a role in differential ASR performance.

\subsection{Aphasia Subgroup Analysis}

While a standard audit might yield results similar to those described in our initial results comparing all aphasia speakers to all control speakers, we instead propose examining aphasia speakers through disaggregated subgroup analyses to account for their heterogeneity. This sentiment is informed by our qualitative interviews with aphasia patients: some participants noted that transcription preferences may vary with rehabilitation stage. For example, in contrast to the general preference for the cleanest possible transcription, a subset of participants earlier in recovery or still undergoing speech therapy expressed interest in less-cleaned transcriptions that could serve as a static record of speech progress that they could observe over time. We proxy these different speech stages via existing medical categorizations of aphasia. Aphasia can be categorized as either \emph{fluent} (speakers can produce sentences but with incorrect words/sounds) or \emph{non-fluent} (speakers struggle with word production and speak in short sentences) \cite{le2023aphasia}, and further subdivided into eight types based on the Boston Diagnostic Aphasia Examination (BDAE)~\cite{goodglass2001bdae}, ranging from Global aphasia (most severe non-fluent) to Anomic aphasia (milder fluent symptoms) \cite{knopman2011regional}. More details are provided in Appendix~\autoref{aphasia-category}. 

\autoref{fig:merged_figure1} shows average WER performance by ASR service for control speakers (from 1,843 audio files) as compared to aphasia speakers by subtype: non-fluent and fluent (from 1,437 and 2,346 audio files, respectively). 
We find that disaggregating aphasia speakers showcases that non-fluent aphasia has significantly worse (higher) average WER relative to fluent aphasia (WER of 0.21 versus 0.13, respectively, with $p<0.001$ for all ASR services). In fact, some fluent aphasia speakers have WER performance much closer to control speakers than to less-fluent aphasia speakers, though both fluent and non-fluent aphasia speakers have significantly worse (higher) average WER relative to the control group ($p<0.001$ for all six ASR services). This finding is robust across WER aggregation method (as defined in \S\ref{sec:wer_def}) and propensity-matched subsets (see Appendix \autoref{fig:groundtruth_comparison_control}).

\begin{figure}[!ht]
    \centering
    \includegraphics[width=0.8\textwidth]{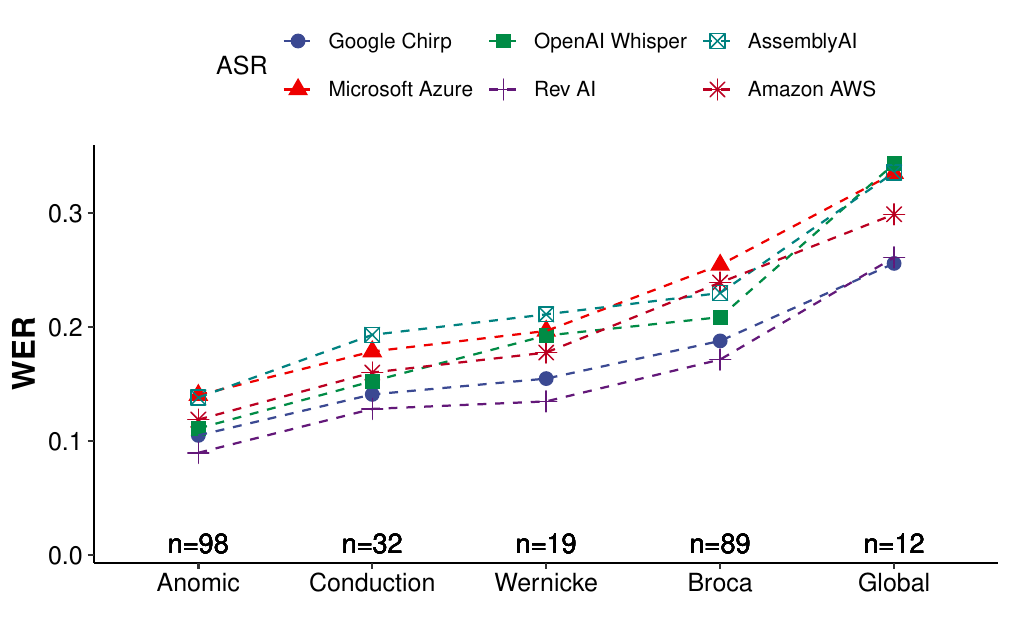}
    \caption{Average WERs for aphasia types are, as expected, worse (higher) for aphasia types that have more severe clinical symptoms: average ASR performance is significantly worse for Global aphasia (WER of 0.31) than for Anomic aphasia (WER of 0.12). Furthermore, performance of each ASR service varies depending on type of aphasia speaker; for example, while OpenAI Whisper overperforms relative to other ASR services for Control, Anomic aphasia, and Conduction aphasia speakers, it underperforms relative to other ASR services for speakers with more severe cases of aphasia. (Note: \textit{n} indicates number of participants from each group, each of whom speaks in multiple audio files---mapping to 1,843 control, 1,454 Anomic, 655 Conduction, 230 Wernicke, 1,058 Broca, and 97 Global aphasia audio files.)} 
    \label{aphasia-type1-wer}
\end{figure}

We further disaggregate ASR performance by more granular aphasia classifications per the BDAE in \autoref{aphasia-type1-wer}. Of the eight categories of aphasia, five are represented in the AphasiaBank data. First, we again find that WER performance can differ significantly within the set of aphasia patients: the Broca and Conduction aphasia groups have significantly higher WER than the Anomic aphasia group ($p<0.001$). Second, we again find that all types of aphasia exhibit worse (higher) average WERs compared to the control group (all pairwise p-values $<0.05$, across all ASR services). Among them, Global aphasia speakers have the highest average WER at 30.5\%, followed by Broca's aphasia at 21.5\%, Wernicke's aphasia at 17.8\%, Conduction aphasia at 15.9\%, and Anomic aphasia at 11.7\%---consistent with the ordering of fluency expected from clinical impressions per Appendix \autoref{aphasia-category}. 

Overall, these findings point towards the need to disaggregate broad groups to better understand heterogeneity within groups, making it imperative that audits report evaluations on more granular subgroups to help inform third-party users about which might be the best ASR service for them to use. This is especially true for audit studies that generate rank orderings of ASR services based on WER performance, such as leaderboards, which are public rankings that compare the performance of different models or services on standardized benchmarks. For example, the Hugging Face Open ASR Leaderboard (\url{https://huggingface.co/spaces/hf-audio/open_asr_leaderboard}) ranks speech recognition models based on their WER across multiple datasets. These rankings (e.g., each vertical column in \autoref{aphasia-type1-wer}) change depending on what subtype of aphasia is being evaluated in the audit. For example, while OpenAI's Whisper performs third-best out of six ASR services on Anomic aphasia (a fluent aphasia with mild symptoms), Whisper performs the worst on Global aphasia (the most severe non-fluent aphasia).

\begin{figure}[!h]

    \centering   \includegraphics[width=0.8\textwidth]{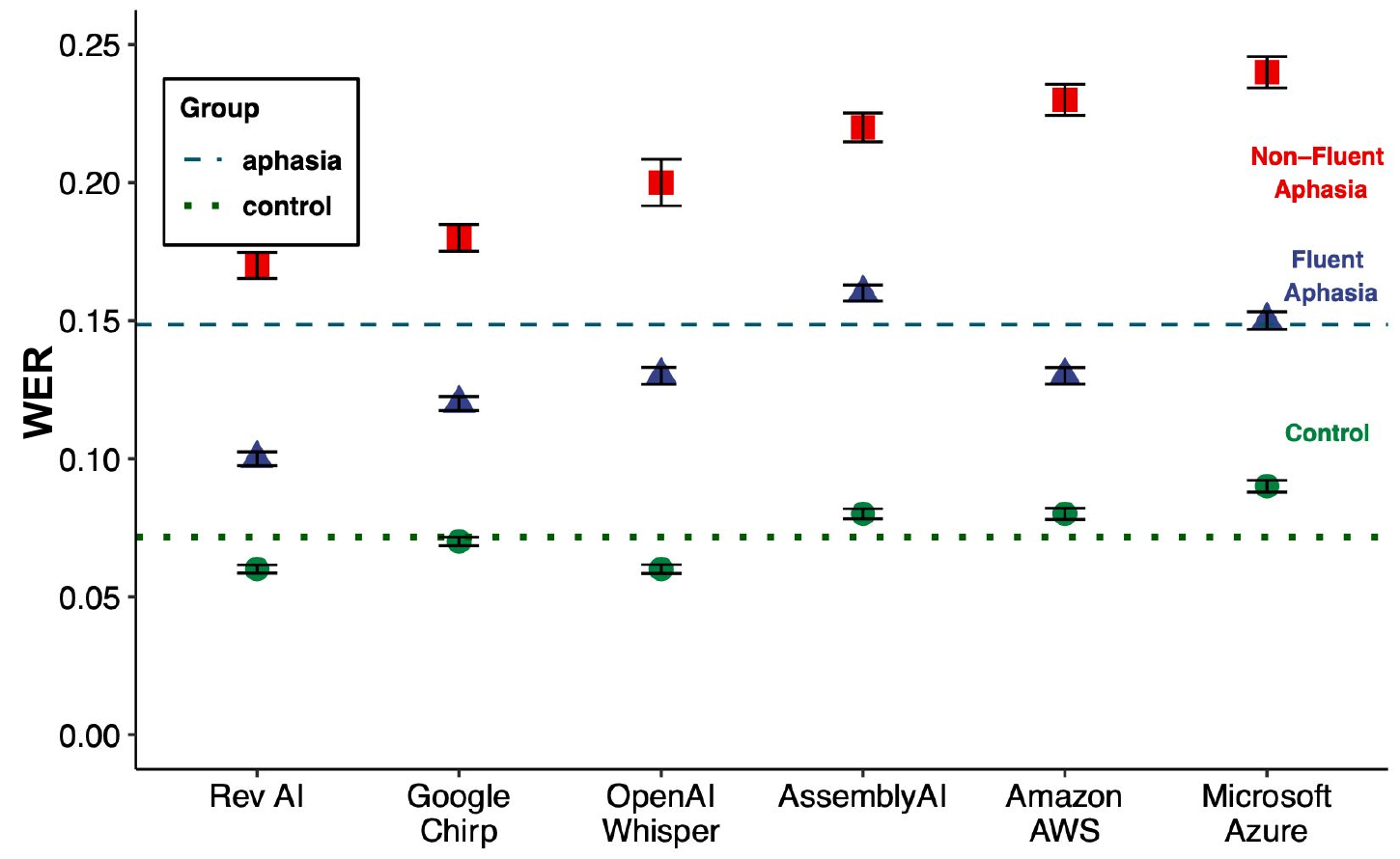}
    \caption{Disaggregating aphasia speakers into fluent and non-fluent speakers showcases the significantly-worse average ASR performance for non-fluent aphasia speakers (red squares, averaging 0.21 WER) relative to fluent aphasia speakers (blue triangles, averaging 0.13 WER). In many cases, ASR performance for fluent aphasia speakers is more comparable to control speakers (green circles, averaging 0.07 WER as represented by the dotted green line) than to non-fluent aphasia speakers. The dashed blue line indicates the average WER across ASR services for all aphasia speakers---the expected granularity of a standard ASR audit's reporting---which obfuscates the disproportionately worse performance for the subset of non-fluent aphasia speakers. Error bars indicate standard errors of the average WER (weighted by word count per \S\ref{sec:wer_def}) for each aphasia type and ASR service.}
\label{fig:merged_figure1}
\end{figure}

\subsection{Other Relevant Covariates that Drive ASR Performance}\label{sec:audiocovariates}

Beyond granular aphasia types, intersectional subgroups also matter: prior work shows ASR services perform worse for Black men and better for white women~\cite{koenecke2020racial}. We analogously find that gender performance gaps widen for more severe aphasia cases (see Appendix Figures \ref{fig:wer_average_gender_unmatched_unweighted} and \ref{fig:wer_average_race_unmatched_unweighted} for analyses on gender and race/ethnicity, respectively, intersected with aphasia type). Additionally, it is imperative to consider audio-level confounders that may not be fully captured by speaker demographics.

We introduce nonvocal duration share and background noise as audio features that are known to directly affect ASR performance and easily computable in Python, but are infrequently included in standard industry audits.
Nonvocal duration share measures the percentage of the number of seconds within the full audio file that does not have a vocal segment present, generally caused by long pauses of silence between utterances. 
This metric serves as a proxy for speech dysfluency, and can be associated with speech impairments~\cite{koenecke2024careless}, age \cite{geng2022speaker}, and emotions \cite{laukka2008nervous}. Nonvocal duration can be estimated using software for Voice Activity Detection~\cite{team2021silero,bredin2020pyannote}; this and other prosodic features more generally have been included in academic studies of ASR performance~\cite{salim2023automatic, johnson2022automatic, goldwater2010words, koenecke2024careless,ris2001assessing}, but are rarely included in industry audits~\cite{assemblyai2024,hughes2023ursa,radford2023robust,deepgram2024}.

In addition to nonvocal duration, background noise represents another critical audio-level confounder. The average background noise in an audio file is approximated by the energy of the overall audio signal. Prior work includes significant methodological development to ensure high ASR performance despite differing levels of background noise~\cite{rodrigues2019analyzing, higuchi2016robust,narayanan2013role}. It is especially important to study ASR performance conditional on background noise levels, since louder background noise tends to correlate both with worse ASR performance, and with lower socioeconomic status~\cite{casey2017race, dale2015socioeconomic,papakyriakopoulos2023augmented}. 

To further examine the effect of various speech features on WER, we fit linear regression models with clustered standard errors at the participant level to isolate the independent effect of aphasia on WER while controlling for these identified confounders.
Clustering on participant level enables us to account for the non-independence of multiple observations (e.g., audio segments) contributed by the same participant. As a robustness check, we also applied linear-mixed effects models---treating participant and audio segment as random effects---to examine whether the effects of our predictors are consistent (see Appendix Table~\ref{tab:mixed_effects_model}). 

To perform regression analysis predicting WER, we calculate both aforementioned metrics for each audio file in the AphasiaBank data (see detailed calculation procedures for these features in \S~\ref{sec:acoustic-features}),
and include them as covariates along with demographic speaker characteristics and other audio file attributes (including total audio file duration in seconds, and total word count in ground truth, since file length also correlates with ASR performance~\cite{siegler1995effects,koenecke2020racial}). The leftmost column of \autoref{tab:merged_regression} lists the coefficients on a linear regression on WER (run on all aphasia and control group audio files), reporting standard errors clustered on individual speakers. As expected, we find that being an aphasia speaker statistically significantly increases the expected WER by 0.07 relative to being a control speaker, all else equal. 

Our demographic findings align with previous research: WER is lower for middle-aged speakers but higher for elderly speakers relative to for young populations \cite{vipperla2010ageing,werner2019automated}, and higher for African American than white speakers~\cite{koenecke2020racial}.
Results remain robust across using propensity-matched datasets and using probit models (Appendix Tables \ref{tab:regression_matched} and \autoref{tab:probit}); using more granular aphasia breakdowns (i.e., binary indicators by subcategory rather than a single binary indicator for aphasia versus control) confirms that more severe cases of aphasia correspond to worse WERs (Appendix Tables \ref{tab:regression_hallucination_fluent}---\ref{tab:mixed_effects_model}).

Importantly, both audio features of interest statistically significantly increase WER: each percentage point increase in nonvocal audio duration corresponds to a 0.002 WER increase, and each unit increase in average background noise energy level increases the WER by 0.07 percentage points. In practice, average energy measures are small units---0.0046 for control speakers and 0.0042 for aphasia speakers---meaning that background noise yields orders of magnitude less impact on WER than being a speaker with aphasia. 
This is consistent with the content of our qualitative interviews: while participants discussed the effect of having aphasia on their ASR usage, no participants independently brought up qualities such as loud background noise being a concern.
Our findings indicate that---while not the primary drivers of WER---audio features such as nonvocal durations and background noise should be included as confounders when isolating effects of aphasia on WER. 

\section{Pitfall 3: Use of a Single Reporting Metric, Word Error Rate (WER)}\label{sec:pitfall-relying_on_wer}

The final issue we highlight is the over-reliance on a single numeric metric---WER---in current auditing practices. While WER is the de facto metric for evaluating ASR model performance \cite{he2011word,huggingface2024leaderboard,hughes2023ursa,assemblyai2024,radford2023robust,deepgram2024}, we argue that it does not fully encapsulate the full range of errors that occur during transcription, which can mask disproportionately severe real-world impact. The argument for introducing suites of evaluation metrics (rather than leaderboards focused on specific metrics) has been made more generally~\cite{wang2024benchmark} for tracking fairness of AI models. For ASR specifically, the case has been made for moving away from only measuring WER because it does not capture severity of errors weighted by semantic importance~\cite{morris2004and}. For example, for the ground truth phrase \emph{``the metrics,''} both transcriptions \emph{``metrics''} and \emph{``the mets''} have a WER of 0.5, but the former transcription preserves the original phrase's meaning more than the latter transcription. 
WER has several limitations: it may not be robust to differences in text standardization, can be averaged across audio files in different ways (per \S \ref{sec:wer_def}), and may not appropriately capture semantic differences or represent the relevant unit level of analysis.

There are several existing metrics that address longtime concerns about the robustness of WER. One concern, as exemplified above, is the lack of weighting by semantic similarity in mistranscriptions. There are many alternative metrics that address this: Word Information Lost (WIL) captures semantic similarity of transcription errors~\cite{morris2004and} (building upon older metrics such as Relative Information Lost (RIL) developed in the 1950s~\cite{miller1955note}); many metrics developed for machine translation, such as METEOR~\cite{banerjee2005meteor} and BERTScore~\cite{zhang2019bertscore}, similarly account for the semantic meaning of words transcribed.

Another concern is whether the unit of a \emph{word} is optimal for evaluation. This may be especially true in non-English languages that are character-based, such as Chinese. We recommend that audits also consider alternative granularities: Character Error Rate (CER), Phoneme Error Rate (PER), or Slot Error Rate (SER) for more granular analysis \cite{wang2003word}, or Sentence Error Rate (SER)~\cite{chou1994minimum} or paraphrase quality~\cite{mirzaei2018exploiting} for less granular evaluation suited to multipurpose ASR tools.

The final concern we highlight---and the focus of this section---regards measuring \emph{hallucinations} in ASR transcriptions. Recent work has pointed to the production of entirely fabricated phrases or sentences in ASR transcriptions by OpenAI's Whisper~\cite{koenecke2024careless,frieske2024hallucinations} that do not appear in the original audio. Hallucinations are markedly different from simple mistranscriptions (which may involve an understandable substitution, such as the earlier ``the mets'' example); they instead involve word additions that are significantly different from the ground truth (for a ground truth of \emph{``...I became ill with a fairly serious strain of viral something''}, Whisper instead transcribed \emph{``...I became ill with a fairly serious strain of viral something, \textbf{but I didn't take any medication, I took Hyperactivated Antibiotics and sometimes I would think that was worse''}})~\cite{koenecke2024careless}. The harm from such hallucinations---i.e., fabricated content that can mislead users in critical applications~\cite{koenecke2024careless}---is not adequately captured by WER (or by its constituent component, the Insertion Rate---the number of word insertions divided by the number of words in the ground truth): as a counterexample, transcribed stutters repeating a word more frequently than the ground truth could similarly yield many insertions, but not represent a hallucination. 

In our qualitative study, we presented distilled examples and high-level results (showcasing hallucinations arising from Whisper) from~\citet{koenecke2024careless} to participants; in response, all participants expressed that hallucinations were a concern. This concern was borne out: reporting by~\citet{burke2024} found that over 7 million doctor visits (from over 30,000 clinicians and 40 health systems) were transcribed by an ASR tool, Nabla, that was built on Whisper. It stands to reason that hallucinations appearing in medical scribe contexts---disproportionately so for patients with aphasia---is a real concern~\cite{koenecke2025perspective}.

To study this quantitatively, we follow prior work in quantifying hallucinations as a binary indicator for whether a hallucination occurs in each audio file, and since there is no consistently high-performing method for identifying hallucinations in ASR transcriptions \cite{choi2025hallucinations}, we manually identify instances as described in the Appendix \S\ref{sec:hallucination-pipeline}. We only identified hallucinations in OpenAI Whisper (finding a hallucination rate lower-bounded at 0.9\%, with 56 confirmed hallucinations among a subset of manually reviewed files---53 of which occurred for aphasia speakers), comparable to the prior literature~\cite{koenecke2024careless}, and found no instances of hallucinations in the other ASR services tested at the time of this study (also consistent with prior work). Experiments on a subset of audio files wherein we perform minor audio editing (such as appending additional seconds of silence to the file) induce more Whisper-hallucinated transcriptions for edited audio files of aphasia speakers versus the control group (85 versus 65 hallucinations, respectively); see Appendix \S \ref{sec:hallucination-results} for details. 

Furthermore, we ran a logistic regression on the hallucination binary, conditioned on demographic and audio file characteristics, in the rightmost column of \autoref{tab:merged_regression}. Similar to findings from Pitfall 2 (analysis of overly-broad user groups), we find that both acoustic metrics of interest (nonvocal audio duration share, and background noise) have positive effects on the likelihood of a Whisper hallucination being transcribed. The nonvocal audio duration share (a proxy for dysfluencies) has a strong significant effect on Whisper hallucinations, as does the speaker having aphasia. These results suggest that certain acoustic characteristics---especially ones disproportionately occurring for aphasia patients---can correspond with higher hallucination rates, underscoring the need for improved ASR services for vulnerable populations who are likely to speak with dysfluencies.

As ASR models quickly develop and fundamentally change as technologies, the suite of relevant metrics should similarly expand---for example, to include hallucination rate as a relevant metric---to ensure that we are auditing for the most relevant errors as ASR technology advances. In addition to hallucination rates, we report our audit on a full suite of additional benchmark metrics in Appendix \autoref{tab:asr_metrics}, finding that the control group yields better ASR performance than the aphasia group across all metrics quantified (BLEU, CER, ROUGE-1/2/L, METEOR, WIL, RIL, and Insertion Rate).

\begin{table}[!htbp]
\centering
\caption{Regression analyses on WER (numeric rate) and Hallucination (binary indicator) with parentheses reporting clustered SE on participant. A linear regression was used to estimate WER conditioning on demographic factors, audio snippet features, and ASR (reference levels are the Control group, Rev AI, white, and less than college education); a binary logistic regression was used to estimate hallucination likelihood among transcriptions from OpenAI's Whisper (the only ASR service yielding hallucination rates at the time of this study).}
\label{tab:merged_regression}
\footnotesize  
\begin{tabular}{@{\extracolsep{0.5pt}}l*{2}{c}}
\toprule
 & \multicolumn{2}{c}{\textit{Dependent variable:}} \\
\cmidrule(lr){2-3}
 & WER & \makecell{Whisper\\Hallucination} \\
\midrule
Aphasia (Binary) & 0.066$^{***}$ (0.003) & 1.508$^{***}$ (0.347) \\
Gender (Female) & $-$0.025$^{***}$ (0.002) & $-$0.522$^{***}$ (0.112) \\
Age & $-$0.005$^{***}$ (0.001) & 0.091$^{*}$ (0.053) \\
Age$^2$ & 0.0001$^{***}$ (0.00001) & $-$0.001$^{*}$ (0.0004) \\
\textless{} 4-year College & 0.024$^{***}$ (0.004) & 0.189 (0.222) \\
4-year college & 0.046$^{***}$ (0.004) & 0.399$^{*}$ (0.206) \\
Post-grad Degree & 0.031$^{***}$ (0.004) & 0.904$^{***}$ (0.192) \\
Race (Other) & 0.008 (0.009) & 0.760$^{**}$ (0.306) \\
Race (African American) & 0.049$^{***}$ (0.006) & 0.282 (0.174) \\
Nonvocal Audio Duration Share (Ratio) & 0.232$^{***}$ (0.009) & 2.127$^{***}$ (0.254) \\
Audio Mean Background Noise & 0.073$^{***}$ (0.007) & 1.059$^{***}$ (0.298) \\
Word Count & 0.0003$^{***}$ (0.00003) & 0.020 (0.013) \\
Total Audio Duration (Seconds) & $-$0.002$^{***}$ (0.0001) & $-$0.117$^{***}$ (0.017) \\
ASR (Google Chirp) & 0.016$^{***}$ (0.004) & --- \\
ASR (Microsoft Azure) & 0.069$^{***}$ (0.004) & --- \\
ASR (OpenAI Whisper) & 0.029$^{***}$ (0.005) & --- \\
ASR (AssemblyAI) & 0.048$^{***}$ (0.004) & --- \\
ASR (Amazon AWS) & 0.046$^{***}$ (0.004) & --- \\
Constant & 0.151$^{***}$ (0.014) & $-$9.333$^{***}$ (1.637) \\
\midrule
\multicolumn{3}{l}{\textit{Note:} $^{*}$p$<$0.1; $^{**}$p$<$0.05; $^{***}$p$<$0.01} \\
\end{tabular}
\end{table}

\section{Discussion}

Given AI systems' rapid societal integration and potential for harm \cite{smuha2021beyond, metcalf2021algorithmic, raji2022fallacy}, AI auditing has become critical, mandated by regulations like the EU AI Act \cite{eu_ai_act} and NYC Local Law 144-21 \cite{nyc_local_law_144} \cite{raji2020closing}. Through our ASR performance case study on aphasia speakers, we identify three actions that practitioners and researchers must take to strengthen auditing practices: (1) practitioners should test multiple text standardization approaches and engage communities to determine preferences rather than imposing nominal standardization assumptions; (2) auditors must disaggregate beyond broad demographics to capture granular subgroup disparities and intersectionally amplified harms; and (3) practitioners should adopt metric suites beyond singular reliance on WER to capture novel errors like hallucinations. 

First, practitioners should test multiple standardization approaches and report performance variance across methods to ensure results are reproducible, robust, and aligned with community needs. Our findings demonstrate that different standardization approaches yield divergent ASR evaluation results. Our aphasia speaker surveys and interviews give insight into their ASR output preferences; subsequent studies should consider other user groups with different speech types and transcription preferences, and auditors should conduct surveys of their target populations. 
Our work has broader implications for text standardization in AI auditing, such as image-to-text models where ground truth captions vary by provider. Addressing standardization challenges requires practitioners' ongoing efforts as technologies and user preferences evolve over time.

Second, auditors must disaggregate analyses to focus on specific demographic subgroups and speech features affecting ASR performance. We measure disparities within aphasia subgroup types and by acoustic features; existing auditing practices may not account for such granular populations. Current approaches adopt a one-size-fits-all methodology, assuming universal speech patterns. While this aligns with anti-discrimination regulation focused on protected categories, such as race, sex, or disability (e.g., the four-fifths rule in U.S. discrimination law \cite{four_fifths_rule}), it ignores fine-grained intersections. We encourage practitioners to incorporate qualitative or mixed-methods approaches to enrich subgroup-level understanding, allowing for deeper insights into lived experiences that can better inform inclusive ASR design and AI auditing. Practitioners must also balance granular data collection with speaker privacy protection.
  
Third, practitioners should adopt metric suites to address the limitation that hallucinations are a distinct class of errors not captured by WER alone. Our findings underscore the need for metric suites measuring semantic transcription differences. Appropriate metrics depend heavily on stakeholders' needs and ASR deployment contexts, requiring continuous refinement as technologies evolve. Since an audit is often the first step toward improving performance and ameliorating biases of ASR services, metrics used in audits must serve this longer-term goal. Using metric suites allows auditors to better pinpoint underlying mechanisms causing worse performance, whether due to model overfitting, label noise, biased training data, data processing, model architecture, or other factors \cite{gekhman2024does, raunak2021curious,koenecke2020racial, bain2023whisperx, gandhi2023distil}. This is especially true as ASR transcription components are replaced with robust end-to-end models for tasks like audio translation~\cite{barrault2023seamlessm4t, alumae2025striving} or summarization, requiring different types of metrics for open-ended evaluation and improvement. 

We recognize the limitations of generalizing from research on clinical populations given data scarcity---our community engagement sample is small, our empirical results derive from a single data source, and we focus on U.S. English. However, we believe our broader framework makes important contributions regarding auditing practices themselves that could be generalizable to people with clinical speech conditions (such as the stuttering or dysphonia communities)~\cite{lea2023user, moore2018whistle, mujtaba2024lost}, and furthermore to other marginalized populations (such as English language learners who may speak with linguistic disfluencies similar to our case study of aphasia patients, or speakers of non-standard dialects)~\cite{wassink2022uneven, harris2024modeling, jahan2025unveiling} in ASR auditing.

Future work can and should seek larger, more diverse samples and examine whether standardization preferences vary by aphasia type and rehabilitation stage. More broadly, our analysis raises but does not fully resolve the question of whether standard cleaning conventions are universally appropriate across different speech communities and ASR deployment contexts where such conventions may inadvertently erase linguistically meaningful features.
Different stakeholders may have distinct preferences; 
future auditing should adopt user-centered frameworks accommodating diverse needs. Improving AI auditing requires continuous adaptation and engagement with stakeholder groups ensuring evaluation metrics remain relevant and comprehensive. 
We encourage future auditing practices to build upon our case study for aphasia patients, and adopt a community-based approach to integrate the diverse preferences of different individuals affected by these technologies.

\section*{Generative AI Usage Statement}
\label{sec:AI Statement}
Generative AI was only used for specific coding tasks (including LaTex table generation and transcript cleaning logic, such as RegEx), and an attempt to classify transcript hallucinations that was not used in the final paper. Generative AI was additionally used for occasional wordsmithing purposes in-text, but was not used for project ideation or the vast majority of writing.

\begin{acks}
We thank AphasiaBank (supported by grant NIH-NIDCD R01-DC008524) for providing the data used in this study, as well as the Boston University Conversation, Health, Art, Technology (C.H.A.T) group for participating in the study.
This work was supported by grants from Mastercard, Inc, the Pulitzer Center and the Cornell Center for Social Sciences. Any views, opinions, findings, and conclusions or recommendations expressed in this material are those of the authors and should not be interpreted as reflecting the views, policies or position, either expressed or implied, of these organizations.
\end{acks}

\bibliographystyle{ACM-Reference-Format}
\bibliography{sample-base}

\appendix

\section{Appendix}
We describe our data filtering, standardization, and matching procedures below, as well as our process for hallucination and WER analyses. Code for reproduction can be found on GitHub (\url{https://github.com/koenecke/auditing_asr_aphasia}).

\subsection{Data Collection and Processing Details}\label{sec:cleaning-details}
\subsubsection{Ground Truth Cleaning and Audio Segmentation}
The raw transcript went through a cleaning process to create a clean ground truth that we refer to when comparing WERs across different APIs. All ground truth cleaning was done programmatically in Python (ver 3.12) using regular expressions. First, all fillers, marked with `\&-’, and all phonological fragments, marked with `\&+’, were coded separately as placeholders. Then, miscellaneous coding such as different types of terminators (e.g., `+...’ for trailing off, `+//’ for self interruption) and utterance level error codes (e.g., `[+ gram]’ for grammatical error, `[+ jar]’ for jargon) were cleaned. As CHAT convention codes repetitions with markers such as `< >’ or `[/]’ depending on the utterance type, the markers were removed to ensure all spoken words, including repetitions, were included in the ground truth. Any word that is marked as unintelligible (`xxx’) in the transcript was marked as `UNK’ in our ground truth and then removed.

Since some audio segments did not contain any content words (as opposed to function words), we filtered the data samples to include only segments with more than three words after removing only filler words.
We segmented the original interview audio files according to the time stamps provided in milliseconds. Since most audio files were in the format of either MP4 videos or MP3 audio, all of the files were converted into mono channel WAV audio files with sampling rate 44.1 kHz and 16-bit PCM using the ffmpeg (version 6.0) \cite{tomar2006converting} tool in Linux. The AudioSegment module from Python’s Pydub (version 0.25.1) \cite{robert2018pydub} library was used for audio segmentation.

To prevent possible timestamp errors (inaccurate timestamps generated by the dataset creators that cause a mismatch between the audio file and the human generated transcript) which can occur when utterances are segmented too shortly and to keep the content consistent, we concatenated all participant utterances that are uttered in one turn (i.e., in between interviewer speech.). This decision was made after multiple timestamp errors were found where the timestamps were not precise enough and caused portions of a word or a phrase to be cut off and not included in the audio file, thus not matching the human-generated transcription. Concatenating to make the snippets longer allows for less room for error in timestamping very short breaks between utterances. For interviews where the interviewer speech was not transcribed, a placeholder was used to mark where the interviewer starts their turn and this was used to segment participant speech. A portion of the interviews was already edited on the raw video level or the raw transcript level from the AphasiaBank to only include participant speech.

Because OpenAI Whisper had a size limit of 25 MB, we decided to enforce a maximum cutoff of 4 minutes to ensure that no audio snippet would be excluded due to ASR service-based size errors. Participant speech was concatenated, but when the concatenated speech block exceeded the 4 minute limit, we looked for the nearest timestamp boundary of more than 1 second and enforced the creation of another block. For a similar reason (RevAI’s minimum size limit of 2 seconds), we discarded any snippets below this range.

When a participant makes a speech error in their utterance that involves wrongly pronounced or misspoken words or phrases, it is the CHAT convention to include the target word or phrase along with the type of the error. These are termed `paraphasia’ or `neologism’, depending on how much of the original term is retained, and phonemic paraphasia is one of the distinguishing features of Aphasia -- hence the need for the extensive transcription. Because the goal of our ground truth is to have an alphabetic representation, we chose to only keeping the orthographic transcription, and thus our main question in cleaning this was to ask whether the erroneous word lacks orthographic representation or not. If the “wrongly” uttered word or phrase has any number of phonemes that are not represented in standard alphabets, this word or phrase is considered unintelligible and marked `UNK’ (e.g., `\textipa{sp\textturnv{}sboz}', `\textipa{\ae{}n\@s{}b\textschwa{}l}'); while the debate on whether what is considered to be wrongly uttered is extremely important, it is not within the scope of our current discussion. Regardless of whether the target word is known or not, we take what is the closest to the spoken word and mark any others as unknown. The same goes for neologisms, where the spoken word is something newly coined by the participant. If the word is transcribable in standard orthography (e.g., `patheticlike’, `wizardess’), these are retained in the ground truth, but if there is a need for a phonetic transcription outside of standard alphabets, the word is marked as `UNK’. Any coding for gestural events (e.g., ‘\&=laughs’ for laughing, ‘\&=hits:table’ for hitting the table) were removed.

Because the presence of an unintelligible word (represented by the `UNK’ token) may be problematic for having the ASR service understand the entirety of the snippet, we decided to remove any snippets that have ‘UNK’ in the ground truth.

After segmentation, some segments were found to incorrectly include interviewer speech or simply had errors in the ground truth. After reviewing the raw interview files, we found that a subset had incorrectly-tagged speaker IDs or time stamps, which were the main causes of the inaccurate segmentation. To automate the identification of such errors, we calculated four metrics to flag these issues (which were manually reviewed to validate the confluence of metrics). First, we used two speaker diarization services (which divide a given audio stream into multiple segments according to different speaker identities) provided by two of the ASR services we experimented with, Microsoft Azure and RevAI. For example, RevAI outputs ``Speaker 0: xxx. Speaker 1:xxx'' to indicate different speakers in one snippet.  Microsoft Azure outputs ``Speaker\_ID''  such as ``Guest-1'', and ``Guest-2'' in a JSON format. Since both services’ results did not completely match and had a relatively low threshold in determining the presence of multiple speakers, we needed additional metrics to identify incorrectly segmented files. Hence, we additionally generated a metric measuring the word count difference between the actual text (ground truth) and the transcripts produced by the ASRs, which flagged whether more than four APIs produced transcripts that significantly exceeded the word count of the ground truth. Next, we generated a string detection metric using regular expressions to capture strings commonly said by an interviewer (based on interviewer scripts or lists of questions to ask the interviewees). The strings we searched for included phrases such as ``tell me'', ``take a look at'', and ``how to make a peanut butter sandwich.'' We manually checked files where metrics were not aligned -- because Microsoft Azure yielded more false negatives (wrongly flagged as having interviewers when in fact not), we manually checked the files where RevAI was flagged FALSE (not including interviewer speech) and Azure was flagged TRUE (including interviewer speech). We also manually checked a random sample of files (3,425 files) where one or two out of the four metrics disagreed with each other. If the manual check did not align with the other metrics, we overrode them via manual check. After all metrics were applied, we were left with 6,039 Aphasia audio segments and 2,418 Control audio segments. After this, we filtered out the audio segments based on our demographic selection criteria (removing files with missing demographic information and restricting to native speakers of English and racial groups that are present in both aphasia and control groups) and audio selection criteria (total duration that have at least 2 seconds and ground truth word count that has at least four words), resulting in 4,342 Aphasia audio segments and 1,843 Control audio segments for statistical analysis.

\subsection{ASR Transcription and Text Processing Pipeline}\label{sec:asr-details}

\subsubsection{ASR System Implementation}

A Python script was written for each service to create a JSON formatted file to include all the transcripts generated from the speech-to-text service. Depending on the service, audio files were transcribed from either a locally stored directory or the service specific cloud stored directory. 

After the initial transcription generation, we noticed that the performance of transcriptions for the default Google and Microsoft models were extremely low with high WER compared to the other four models (0.57 WER for Google, 0.79 for Microsoft). Because the Google default model often generated empty transcriptions, we instead re-ran Google experiments using different enhanced models (Long and Telephony) and a newly built model (Chirp), bringing Google’s performance up to par with the other services. Because Microsoft Azure performed poorly in the cases where there were longer silences in the audio files---resulting in transcriptions stopping before the end of the audio file due---we re-ran experiments with the Azure Continuous model that ensures the transcription service continues unless told otherwise, instead of automatically assuming silence indicates the end of an audio stream. The Azure Continuous model helped improve the WER, bringing down the WER to a comparable rate to the other ASR services, and errors caused by abrupt endings of transcriptions did not persist. There were audio snippets that failed to be transcribed with any output from ASR services: Google Long failed on 62 files, Google Telephony failed on 77 files, AssemblyAI failed on 75 files, RevAI failed on 29 files, Whisper failed on 12 files, Google Chirp failed on 113 files, Azure failed on 38 files, and AWS failed on 9 files. There were more audio snippets from the aphasia group than the control group that ASR services failed to transcribe, regardless of ASR-induced size limits. We re-ran these files through each ASR for a second time (November 1st, 2023) to confirm that they remained unable to be transcribed. These failed transcriptions were treated as empty strings for WER calculations (yielding a 100\% WER).

\subsubsection{Text Standardization}

Since different ASR services treat certain words in different ways, we programmatically cleaned and standardized the ASR-generated transcripts by using a Python script that we share on a public repository. Our cleaning process mainly uses the whisper\_normalizer (version 0.0.2, released March 22, 2023) package built with the release of OpenAI Whisper. 

Before standardizing the transcript text with the whisper\_normalizer, we took several programmatic steps that were not included in the whisper\_normalizer.  We first standardized anonymization of all markers of first names and last names in the ground truth due to the inconsistency of formatting in AphasiaBank data. For example, ``FirstnameX'' and ``Firstname'' were all converted into ``firstname''. Then for snippets that contained these names, we standardized the first names and last names in ASR transcripts by converting specific names into ``firstname” or ``lastname” (e.g., ``Connie'' becomes ``firstname''). Then we removed words that indicate diarization (speaker number and time stamps) from the transcripts. These types of words in the form like ``Speaker 0 XX:XX”  are only present in RevAI transcripts among our data. RevAI transcripts also include words that describe the speaker’s movement such as ``[laughs]''; we do not use any additional programmatic code to remove these words in brackets because the whisper\_normalizer removes them as mentioned above. Following this step, we standardized time-related text (e.g., ``12:00'' was converted to ``12 o’clock'') to avoid inducing errors from whisper\_normalizer, which converts ``12:0'' into ``12 0'' without capturing the time component. Then, we standardized spellings of words by converting informal or colloquial terms to formal words, abbreviations to full forms, compounding words to separated words, etc. following prior work~\cite{koenecke2024careless,Zhao2024}.

After using the whisper\_normalizer, we took an additional programmatic step to remove punctuation again because the whisper\_normalizer does not remove punctuation within digits like ``5.30'' (five thirty). For numeric strings, we insert space between digits due to the inconsistent removal of space between digits in whisper\_normalizer. For example, if a participant counts numbers in a sequence ``one two 3 4 12345'', the final output from our cleaning is ``1 2 3 4 1 2 3 4 5'' (whereas the whisper\_normalizer standardizes it as ``12 3 4 12345''). Finally, we remove additional filler words not removed by the whisper\_normalizer (`hmhm', `uhhuh', `emmm', `huh', `umm', `ugh', `hm', `uhuh', `eh', `uhh', `mmhmm').

\subsection{Acoustic Feature Engineering}\label{sec:acoustic-features}

We measured the total duration and non-vocal duration for each audio snippet using the Silero voice activity detection (VAD) package \cite{team2021silero} (retrieved on February 2024). VAD algorithms identify segments in audio files where only background noise or silence is present, distinguishing these from parts containing human voice. For our analysis, we computed the duration of vocal segments, which was then subtracted from the total duration of the audio file to yield the non-vocal duration. The average non-vocal percentage within an audio file was 30.44\% for aphasia speakers and 18.22\% for control speakers, indicating a substantial difference in speech patterns between the two groups. 

We evaluated three different VAD packages: Silero \cite{team2021silero} (retrieved on February 2024), Pyannote \cite{bredin2020pyannote} (ver 3.3.2) and WebRTC \cite{py_webrtcvad} (ver 2.0.10). We examined the performance of each package on our specific dataset to determine the most appropriate one for our analysis. Pyannote and Silero showed a high correlation of 0.944 in their detected non-vocal durations. In contrast, WebRTC showed lower correlations with both Silero (0.883) and Pyannote (0.896). This discrepancy may be attributed to WebRTC's requirement for audio files to be formatted at a specific sampling rate, necessitating resampling of our audio files, which could have led to fluctuations in audio quality and specifications. Based on these results and additional qualitative analysis of the detection accuracy, we selected Silero as our primary VAD tool for the analyses reported in the main text.

Background noise was calculated using the Librosa package \cite{mcfee2024librosa} (ver 0.10.2), a widely used Python package for audio analysis. We calculate the energy of the overall audio signal using the Root Mean Square (RMS), with a frame size of 2048 samples and a hop size of 512, which are the default settings in Librosa for energy-based calculations. RMS energy provides a measure of the signal's amplitude, which we used to identify low-energy segments as potential background noise. Segments with RMS energy below a threshold of 0.01 were presumed to represent background noise. For each identified segment, the average energy level was calculated to characterize the background noise for each audio file. The aphasia group exhibited a mean background noise level of RMS background noise level of 0.0043 (-47.32 dB), whereas the control group had a mean RMS background noise level of 0.0048 (-46.36 dB), which is significantly louder though---as with aphasia speakers' background noise levels---still softer than the threshold of human hearing ($t = -12.44$, $p < 0.001$).

\subsection{Statistical Analysis Details}\label{sec:stats-details}
\input{tables/demographic_table}

\subsubsection{Propensity Score Matching Details}
Distributions of speaker and audio file characteristics pre- and post-matching are presented in \autoref{tab:pre-post-matching-count}. Standardized mean differences for covariates before and after propensity matching are visualized in \autoref{fig:loveplot}. To obtain balanced samples across fluent aphasia, nonfluent aphasia, and control groups, we implemented common referent matching with the control group as the common referent group ~\cite{rassen2011simultaneously}, resulting in a total of 1,184 control, 1,633 fluent aphasia, and 1,256 non-fluent aphasia snippets. Specifically, we matched on demographic attributes including gender, race, age, and education levels. Note that to improve propensity matching, we combined Hispanic/Latino and Asian with the `Other' category due to the low number of counts in these categories. 
\input{tables/distributional_balance_table}

\input{tables/love_plot}

\subsubsection{Statistical Tests for System Comparison and Subgroup Comparison}\label{sec:statstests}

Results from the Kruskal-Wallis tests ($p<0.001$ for all six ASR services) reveal that both non-fluent aphasia and fluent aphasia have significantly worse (higher) WER relative to control group. Non-fluent aphasia has significantly worse (higher) average WER relative to fluent aphasia (21\% versus 13\%, respectively, with nonparametric pairwise tests $p<0.001$ for all ASR services).

\subsection{Baseline ASR Performance Results}\label{sec:baseline-results}

\begin{figure}[!h]
    \centering
    \includegraphics[width=\textwidth]{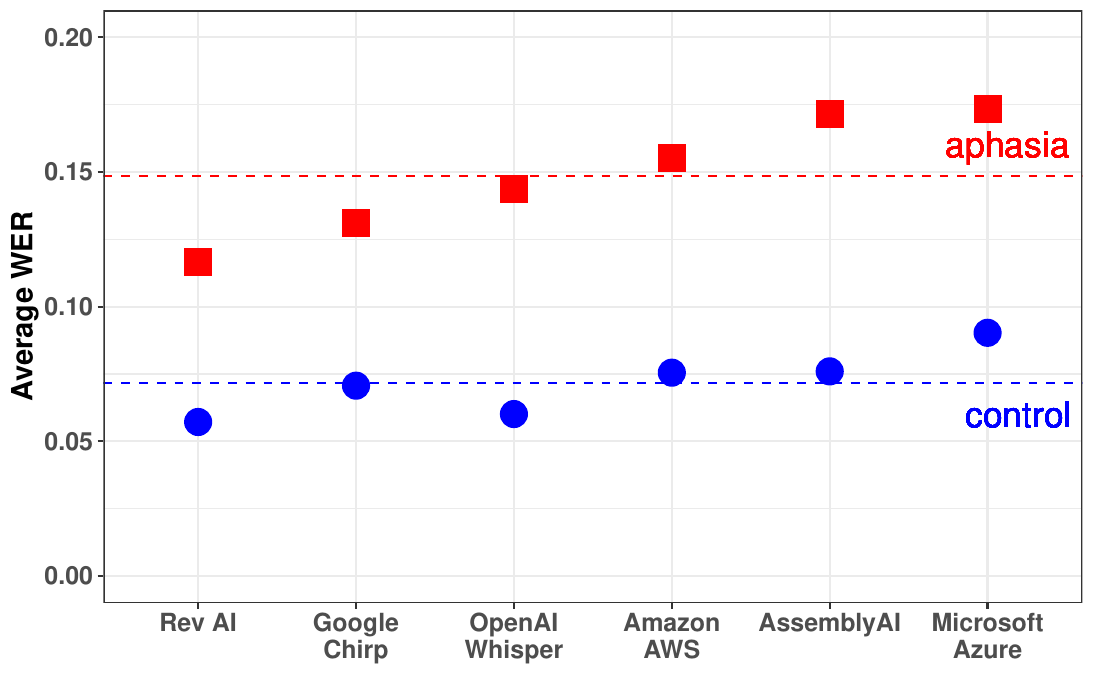}
    \caption{On a total sample of 4,342 aphasia and 1,843 control audio files, we find that WER is consistently worse for aphasia speakers than for control speakers (on average, 0.15 versus 0.07, respectively). There is also variance in performance of ASR systems, with Rev AI performing best and Microsoft Azure performing worst for aphasia speakers in the AphasiaBank corpus. The difference in WER between matched aphasia and control speakers ranges from 6 to 10 percentage points across ASR systems. Results are comparable when using a weighted averaging scheme for WER aggregations per \autoref{sec:wer_def}, which yields a difference in WER between aphasia and control speakers between 5 to 8 percentage points.}
    \label{wer_unmatched_weighted}
\end{figure}

\autoref{wer_unmatched_weighted}, \autoref{tab:wer_weighted_unweighted}, \autoref{tab:wer_test_results} support the baseline performance comparison in Section~\ref{sec2.1}.

\begin{table}[!h]
\centering
\caption{A robust suite of WER calculations demonstrating differences across ASR systems for control and aphasia speakers when using different approaches to averaging WER across audio files (weighted or unweighted by ground truth word count per \autoref{sec:wer_def}) and datasets (unmatched full dataset or matched subset of audio files). Regardless of method, all ASR services underperform on aphasia speakers relative to the control group.}\label{tab:wer_weighted_unweighted}
\renewcommand{\arraystretch}{1.5}
\begin{tabular}{|c|c|c c|c c|c c|c c|}
\hline
\multirow{2}{*}{\textbf{ASR Service}} & \multirow{2}{*}{\textbf{Group}} & \multicolumn{2}{c|}{\textbf{Weighted}} & \multicolumn{2}{c|}{\textbf{Unweighted}} \\ 
&&\textbf{Matched} & \textbf{Unmatched} & \textbf{Matched} & \textbf{Unmatched} \\ \hline
\multirow{2}{*}{Rev AI}  
& Control    & 0.07   & 0.06  & 0.12  & 0.10 \\ 
 & Aphasia    & 0.09   & 0.12   & 0.14  & 0.19   \\
\hline
\multirow{2}{*}{Google Chirp}        
& Control & 0.09 & 0.07  & 0.13 & 0.11\\ 
& Aphasia   & 0.10 & 0.13 & 0.16& 0.21\\ \hline
\multirow{2}{*}{OpenAI Whisper}  
& Control  & 0.08  & 0.06  & 0.12  & 0.10 \\ 
& Aphasia   & 0.11 & 0.14 & 0.18  & 0.23  \\
\hline
\multirow{2}{*}{Amazon AWS}         
& Control& 0.09 & 0.08& 0.15 & 0.12 \\ 
  & Aphasia & 0.12 & 0.16 & 0.19  & 0.24 \\ \hline
  \multirow{2}{*}{Microsoft Azure}     
& Control& 0.11 & 0.09 & 0.18 & 0.15  \\ 
 & Aphasia  & 0.13  & 0.17  & 0.20  & 0.26 \\
 \hline
 \multirow{2}{*}{AssemblyAI}          
& Control  & 0.09 & 0.08 & 0.16  & 0.13 \\ 
 & Aphasia  & 0.14  & 0.17  & 0.20 & 0.24\\ \hline
\end{tabular}
\end{table}

\begin{table}[!h]
\centering
\caption{Mann-Whitney U test results for WER between aphasia and control speakers across six ASR systems. Both matched subset and unmatched samples yielded statistically significant difference in WER between aphasia and control speakers across all six ASR systems.}
\renewcommand{\arraystretch}{1.5}
\begin{tabular}{|l|cc|cc|}
\hline
\textbf{ASR System} & \multicolumn{2}{c|}{\textbf{Matched}} & \multicolumn{2}{c|}{\textbf{Unmatched}} \\
             & \textbf{U}     & \textbf{p-value}     & \textbf{U}      & \textbf{p-value}     \\
\hline
OpenAI Whisper   & 622893.0    & $<0.001$    & 2941098.0    &  $<0.001$\\
Google Chirp     & 692744.5    &  $<0.001$    & 	3180047.5    &  $<0.001$ \\
Rev AI           & 699047.0    & $<0.001$    & 3184226.0    &  $<0.001$ \\
Amazon AWS       & 639525.0    &  $<0.001$    & 2830438.5    &  $<0.001$\\
Microsoft Azure  & 673580.0    &  $<0.001$    & 2879604.5    &  $<0.001$ \\
AssemblyAI       & 614832.0    &  $<0.001$    & 	2753710.5 & $<0.001$ \\
\hline
\end{tabular}
\label{tab:wer_test_results}
\end{table}

\subsection{Text Standardization Robustness Analysis}\label{sec:standardization-robustness}

\begin{figure}[!h]
    \centering
    \includegraphics[width=\linewidth]{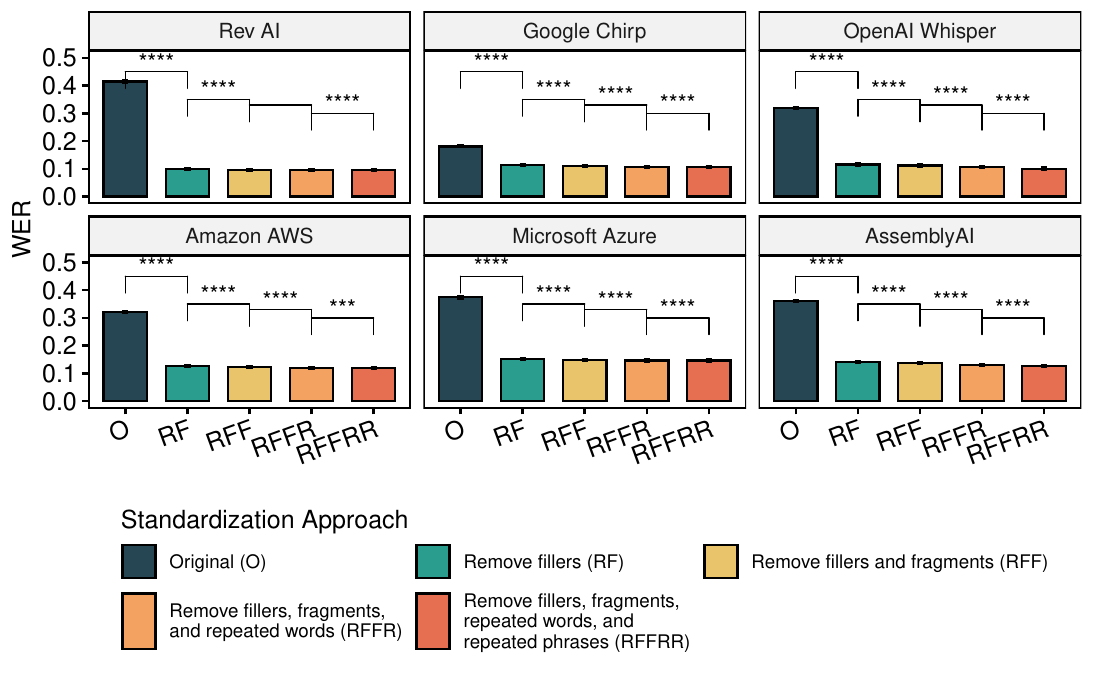}
    \caption{On a total sample of 1,843 control group audio files, we find a statistically significant reduction in WER when additionally removing repeated words for all ASR services except for Rev AI. Removing repeated phrases results in a statistically significant, but small magnitude reduction in WER for all ASR services.  (Significance symbol:$ *: p<0.05; **: p <= 0.01;***: p <= 0.001 ;****: p <= 0.0001$)}
    \label{fig:groundtruth_comparison_control}
\end{figure}

\autoref{fig:groundtruth_comparison_control} and \autoref{tab:wer_standardization} support the discussion in Section~\ref{sec:pitfall-standardization} examining how different text cleaning approaches affect WER. For both aphasia and control files across all six ASR services, there were statistically significant differences in WER across different standardization approaches ($p<0.001$). For control snippets, removing fragments and repeated words did not yield statistically significant differences in WERs across four ASRs: Rev AI, Google Chirp, Amazon AWS, and Microsoft Azure. For OpenAI Whisper and AssemblyAI, removing repeated words yielded a statistically significant reduction in WER.

\input{tables/kruskal_wallis_result}

\begin{table}[!h]
\caption{Average WER across different standardization approaches for different ASR services: average WERs are within 3 percentage points of each other among all standardization approaches. }\label{tab:wer_standardization}
\centering

\begin{tabular}{p{2.4cm}lp{1.2cm}p{1.5cm}p{2.1cm}p{2.1cm}p{2.5cm}}
  \hline
  \multirow{2}{*}{ASR} & \multirow{2}{*}{Group} & \multicolumn{5}{c}{WER} \\ 
  \cmidrule{3-7}
  & & Original (O) & Remove fillers (RF) & Remove fillers and fragments (RFF) & Remove fillers, fragments, and repeated words (RFFR) & Remove fillers, fragments, repeated words, and repeated phrases  (RFFRR) \\ 
  \hline
  \multirow{2}{*}{Google Chirp} & Aphasia & 0.32 & 0.23 & 0.20 & 0.20 & 0.21 \\ 
   & Control & 0.18 & 0.11 & 0.11 & 0.11 & 0.11 \\ 
  \hline
  \multirow{2}{*}{Microsoft Azure} & Aphasia & 0.64 & 0.28 & 0.26 & 0.26 & 0.26 \\ 
   & Control & 0.37 & 0.15 & 0.15 & 0.15 & 0.15 \\ 
  \hline
  \multirow{2}{*}{OpenAI Whisper} & Aphasia & 0.57 & 0.26 & 0.25 & 0.24 & 0.23 \\ 
   & Control & 0.32 & 0.12 & 0.11 & 0.11 & 0.10 \\ 
  \hline
  \multirow{2}{*}{Rev AI} & Aphasia & 0.78 & 0.20 & 0.18 & 0.19 & 0.19 \\ 
   & Control & 0.42 & 0.10 & 0.10 & 0.10 & 0.10 \\ 
  \hline
  \multirow{2}{*}{AssemblyAI} & Aphasia & 0.61 & 0.30 & 0.27 & 0.25 & 0.24 \\ 
   & Control & 0.36 & 0.14 & 0.14 & 0.13 & 0.13 \\ 
  \hline
  \multirow{2}{*}{Amazon AWS} & Aphasia & 0.58 & 0.25 & 0.24 & 0.24 & 0.24 \\ 
   & Control & 0.32 & 0.13 & 0.12 & 0.12 & 0.12 \\ 
  \hline
\end{tabular}

\end{table}

\input{tables/mapsswe_test_result_aphasia}

\newpage

\subsection{Aphasia Classifications and Demographic Disaggregation}

This section provides reference material on aphasia types (\autoref{sec:pitfall-demographic}) and disaggregated performance analyses by demographic subgroups.

\subsubsection{Performance by Aphasia Severity}\label{sec:aphasia-category}
\begin{figure}[!h]
    \centering
    \includegraphics[width=\textwidth]{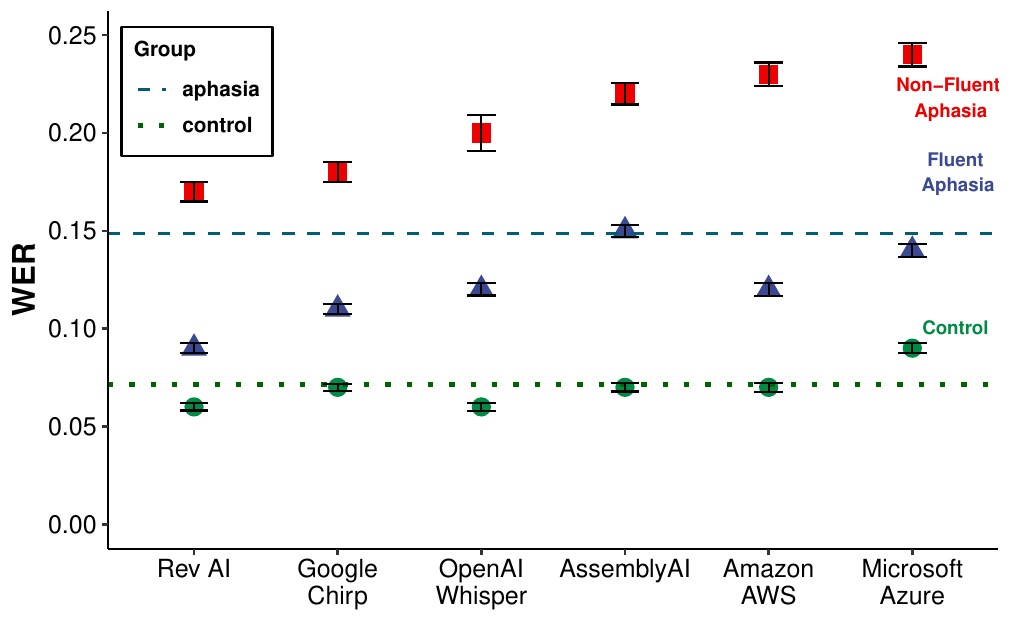}
    \caption{On a total sample of 1,184 control, 1,633 fluent aphasia, and 1,256 non-fluent aphasia audio files, we find that WER across all six ASR systems is worse for non-fluent aphasia and fluent aphasia speakers than for control speakers (averaging 0.21 vs. 0.12 vs. 0.07, respectively) when using a weighted WER average per \autoref{sec:wer_def} on matched subsamples. Additionally, unweighted WER for non-fluent aphasia speakers is notably worse than for control speakers across all six ASR systems. There is also considerable variation in performance across ASR systems for each group: for both fluent and non-fluent aphasia speakers RevAI performs best, while for fluent aphasia speakers AssemblyAI performs the worst, and for non-fluent aphasia speakers Microsoft Azure performs the worst. The error bars represent standard errors of the average WER, indicating low variability of average WER between samples. }
    \label{fig:fluency-matched-weighted}
\end{figure}

\begin{figure}[!b]
    \centering
    \includegraphics[width=\textwidth]{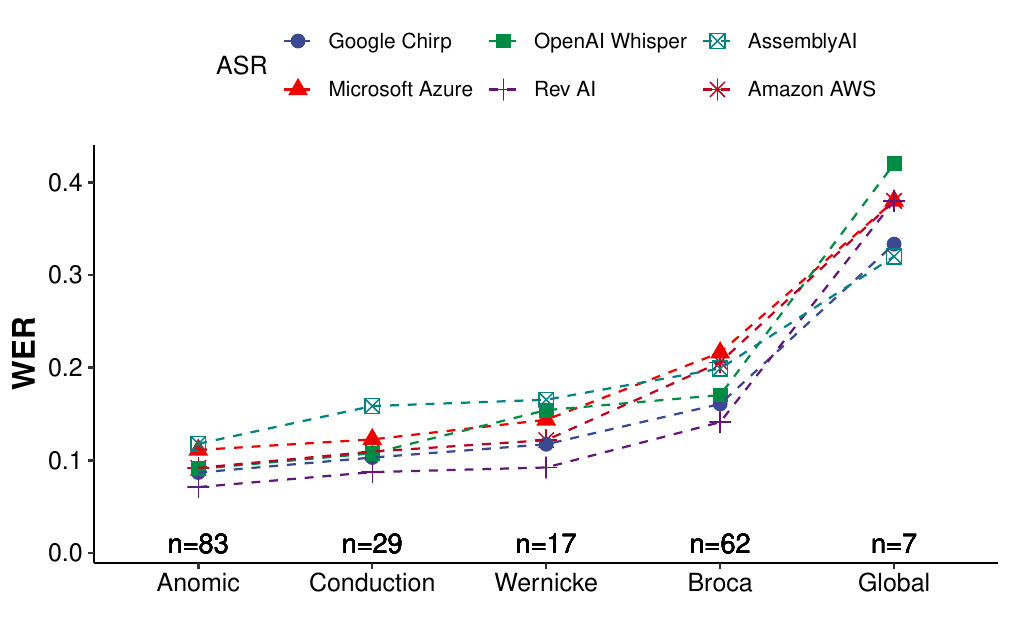}
    \caption{Across a matched subsample of 1,224 control group audio files, and 497 Anomic, 164 Conduction, 63 Wernicke, 223 Broca, and 10 Global aphasia audio files, we find that ASR services yield significantly different WERs across different clinical impressions of aphasia as compared to the control group (\textit{n} indicates number of participants from each group, each of whom may speak in multiple audio files). The average WERs for aphasia types are, as expected, worse (higher) for aphasia types that have more severe clinical symptoms: average ASR performance is significantly worse for Global aphasia (WER of 0.28) than for Anomic aphasia (WER of 0.13). Furthermore, performance of each ASR service varies depending on type of aphasia speaker; for example, while RevAI overperforms relative to other ASR services for Control, Anomic aphasia, and Conduction aphasia speakers, it underperforms relative to other ASR services for speakers with more severe cases of aphasia.}
    \label{fig:boston-matched-weighted}
\end{figure}

Speech capabilities vary across different aphasia types; we list these differences in \autoref{aphasia-category}.
\autoref{fig:fluency-matched-weighted} and \autoref{fig:boston-matched-weighted} disaggregate ASR performance by aphasia fluency (fluent vs. non-fluent) and by the five specific aphasia types represented in our dataset, using propensity-matched samples to control for demographic confounders.

\begin{table}[htbp]
\caption{Speech Capabilities of Individuals with Different Aphasia Types, ordered from most to least severe based on the BDAE classification. ``Speech fluent?'' indicates whether the individual can produce fluent speech. ``Can comprehend?'' refers to the ability to understand spoken language. ``Can repeat?'' indicates whether the individual can repeat words or phrases when prompted.}\label{aphasia-category}
\begin{tabular*}{\textwidth}{@{\extracolsep\fill}lccc}
\toprule%
Aphasia Types & Speech fluent? & Can comprehend? & Can repeat?\\
\midrule
Global & & & \\
Mixed Transcortical & & & \checkmark \\
Broca's & & \checkmark & \\
Wernicke's & \checkmark & & \\
Transcortical Motor & & \checkmark & \checkmark \\
Transcortical Sensory & \checkmark & & \checkmark \\
Conduction & \checkmark & \checkmark & \\
Anomic & \checkmark & \checkmark & \checkmark \\
\end{tabular*}
\end{table}

\subsubsection{Performance by Gender and Race}
\begin{figure}[!h]
    \centering
    \includegraphics[width=0.8\linewidth]{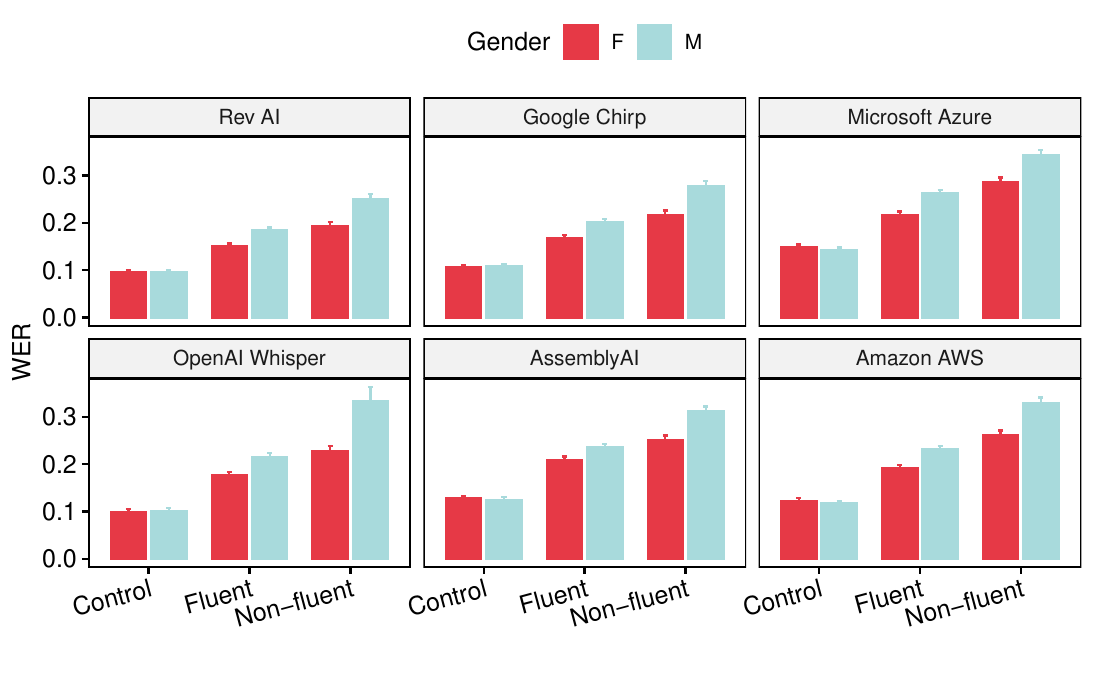}
    \caption{On a total sample of 1,843 control, 2,393 fluent aphasia, and 1,392 non-fluent aphasia audio files, we find that WER is consistently worse for non-fluent aphasia speakers compared to fluent aphasia speakers. A more granular analysis of WER by gender reveals that WER is consistently worse for male (M) aphasia speakers compared to female (F) aphasia speakers across all ASR services, and the difference is compounded for speakers with more severe aphasia (i.e., the gender performance gap is more severe for non-fluent aphasia speakers than for fluent aphasia speakers). There were no non-binary speakers in our data sample.} \label{fig:wer_average_gender_unmatched_unweighted}
\end{figure}

\begin{figure}[h!]
        \centering
        \includegraphics[width=0.8\linewidth]{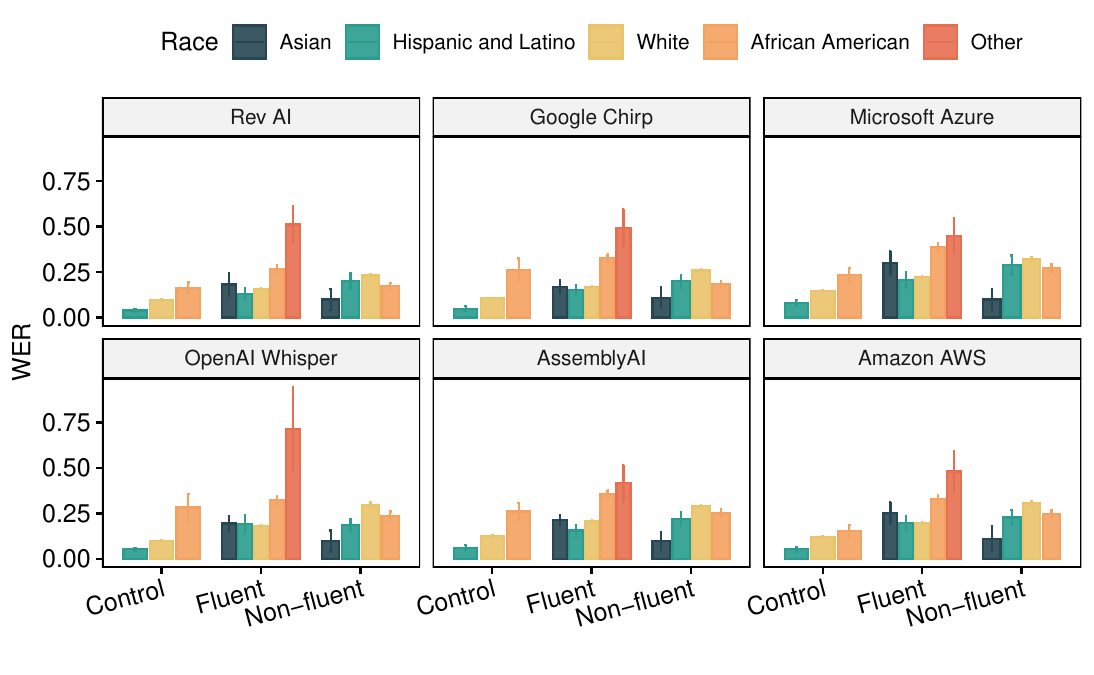}
        \caption{On a total sample of 1,843 control, 2,393 fluent aphasia, and 1,392 non-fluent aphasia audio files, a granular analysis of WER by participants' racial ethnicity reveals that WER is worse for African American speakers than White speakers in both control and fluent aphasia groups across all ASR services.  Note that our sample consists of more White participants (control: 207; fluent aphasia:142; non-fluent:110) than participants from Asian (control:0; fluent aphasia:1;non-fluent:1), Hispanic and Latino (control:5; fluent aphasia:3;non-fluent:3), African American (control:4; fluent aphasia:13;non-fluent:14), and other racial ethnicities (control:0; fluent aphasia:1;non-fluent:0).  }\label{fig:wer_average_race_unmatched_unweighted}
        
    \end{figure}

\autoref{fig:wer_average_gender_unmatched_unweighted} and \autoref{fig:wer_average_race_unmatched_unweighted} present WER disaggregated by gender and race/ethnicity within aphasia subgroups.

\subsubsection{Regression Model Specifications}

\autoref{tab:regression_matched}, \autoref{tab:probit}, \autoref{tab:regression_hallucination_fluent} and \autoref{tab:regression_hallucination_category} present full regression results examining the effect of aphasia status, demographic factors, and acoustic features on WER and hallucination outcomes. In addition, we include mixed effects modeling results in \autoref{tab:mixed_effects_model}.

\begin{table}[h!] \centering 
  \caption{Regression analysis of WER for the matched data by participant demographic, audio segment features, and ASR with clustered SE on participant (reference levels are the control group, Rev AI, white, less than college education).} 
  \label{tab:regression_matched}
 \begin{tabular}
 {@{\extracolsep{0.05pt}}lcc} 
\\[-1.8ex]\hline 
\hline \\[-1.8ex] 
 & \multicolumn{2}{c}{\textit{Dependent variable:}} \\ 
\cline{2-3} 
\\[-1.8ex] & \multicolumn{2}{c}{WER} \\ 
\\[-1.8ex] & \multicolumn{1}{c}{(1)} & \multicolumn{1}{c}{(2)}\\ 
\hline \\[-1.8ex] 
 Aphasia & 0.050$^{***}$ (0.004) & 0.045$^{***}$ (0.004) \\ 
 Gender (Female) & -0.032$^{***}$ (0.004) & -0.025$^{***}$ (0.003) \\ 
  Age & -0.007$^{***}$ (0.001) & -0.007$^{***}$ (0.001) \\ 
  Age$^2$ & 0.0001$^{***}$ (0.00001) & 0.0001$^{***}$ (0.00001) \\ 
  2-year College or dropped Out of 4-year College & 0.016$^{***}$ (0.005) & 0.022$^{***}$ (0.005) \\ 
  4-year college & 0.028$^{***}$ (0.005) & 0.035$^{***}$ (0.005) \\ 
  Post-grad Degree & 0.027$^{***}$ (0.005) & 0.034$^{***}$ (0.005) \\ 
  Race (Other) & 0.068$^{***}$ (0.015) & 0.039$^{***}$ (0.014) \\ 
  Race (African American) & 0.144$^{***}$ (0.019) & 0.096$^{***}$ (0.017) \\ 
  Mean Background Noise (Audio) &  & 0.051$^{***}$ (0.008) \\ 
  Word Count &  & -0.0002$^{***}$ (0.0001) \\ 
  Total Audio Duration (Seconds) &  & -0.001$^{***}$ (0.0001) \\ 
  Non-vocal Audio Duration Share (\%) &  & 0.235$^{***}$ (0.010) \\ 
  ASR (Google Chirp) & 0.013$^{**}$ (0.005) & 0.013$^{***}$ (0.005) \\ 
  ASR (Microsoft Azure) & 0.060$^{***}$ (0.006) & 0.060$^{***}$ (0.005) \\ 
  ASR (OpenAI Whisper) & 0.019$^{***}$ (0.006) & 0.019$^{***}$ (0.006) \\ 
  ASR (AssemblyAI) & 0.047$^{***}$ (0.006) & 0.047$^{***}$ (0.005) \\ 
  ASR (Amazon AWS) & 0.037$^{***}$ (0.006) & 0.037$^{***}$ (0.005) \\ 
  Constant & 0.282$^{***}$ (0.016) & 0.242$^{***}$ (0.015) \\ 
 \hline \\[-1.8ex] 
Observations & \multicolumn{1}{c}{14,688} & \multicolumn{1}{c}{14,688} \\ 
R$^{2}$ & \multicolumn{1}{c}{0.045} & \multicolumn{1}{c}{0.168} \\ 
Adjusted R$^{2}$ & \multicolumn{1}{c}{0.044} & \multicolumn{1}{c}{0.167} \\ 
Residual Std. Error & \multicolumn{1}{c}{0.208 (df = 14673)} & \multicolumn{1}{c}{0.194 (df = 14669)} \\ 
F Statistic & \multicolumn{1}{c}{49.754$^{***}$ (df = 14; 14673)} & \multicolumn{1}{c}{164.204$^{***}$ (df = 18; 14669)} \\ 
\hline 
\hline \\[-1.8ex] 
\textit{Note:}  & \multicolumn{2}{r}{$^{*}$p$<$0.1; $^{**}$p$<$0.05; $^{***}$p$<$0.01} \\ 
\end{tabular} 
\end{table}

\begin{table}[!h] \centering 
  \caption{Probit Model of WER for unmatched data by participant demographic, audio segment features, and ASR with clustered SE on participant (reference levels are the control group, Rev AI, white, less than college education).} 
  \label{tab:probit} 
\begin{tabular}{@{\extracolsep{5pt}}lc} 
\\[-1.8ex]\hline 
\hline \\[-1.8ex] 
 & \multicolumn{1}{c}{\textit{Estimates (Standard Errors) }} \\ 
\\[-1.8ex] &   \\ 
\hline \\[-1.8ex] 
Aphasia & 0.171$^{***}$ (0.014) \\ 
Gender (Female) & $-$0.083$^{***}$ (0.012) \\ 
  Age & $-$0.028$^{***}$ (0.002) \\ 
  Age$^2$ & 0.0002$^{***}$ (0.00002) \\ 
2-year College or dropped Out of 4-year College & 0.108$^{***}$ (0.019) \\ 
4-year college & 0.165$^{***}$ (0.017) \\ 
Post-grad Degree & 0.150$^{***}$ (0.018) \\ 
Race (Other) & 0.128$^{***}$  (0.042) \\ 
Race (African American) & 0.257$^{***}$ (0.038) \\ 
Nonvocal Audio Duration Share (\%) & 0.753$^{***}$ (0.029) \\ 
Audio Mean Background Noise & 0.233$^{***}$ (0.032) \\ 
Word Count & $-$0.002$^{***}$ (0.0003) \\ 
Total Audio Duration (Seconds) & $-$0.005$^{***}$ (0.001) \\ 
  ASR (Google Chirp) & 0.072$^{***}$ (0.021) \\ 
  ASR (Microsoft Azure) & 0.267$^{***}$ (0.020) \\ 
  ASR (OpenAI Whisper) & 0.053$^{**}$ (0.021) \\ 
  ASR (AssemblyAI) & 0.208$^{***}$ (0.021) \\ 
  ASR (Amazon AWS) & 0.160$^{***}$ (0.020) \\ 
  Constant & $-$0.669$^{***}$ (0.058) \\ 
 \hline \\[-1.8ex] 
\hline 
\hline \\[-1.8ex] 
\textit{Note:}  & \multicolumn{1}{r}{$^{*}$p$<$0.1; $^{**}$p$<$0.05; $^{***}$p$<$0.01} \\ 
\end{tabular} 
\end{table}

\begin{table}[!h] \centering 
  \caption{Regression analysis of WER and hallucination by aphasia fluency (fluent vs. non-fluent), participant demographic, audio segment features, and ASR with clustered SE on participant (reference levels are the control group, Rev AI, white, less than college education. } 
  \label{tab:regression_hallucination_fluent} 
\begin{tabular}{@{\extracolsep{5pt}}lcc} 
\\[-1.8ex]\hline 
\hline \\[-1.8ex] 
 & \multicolumn{2}{c}{\textit{Dependent variable:}} \\ 
\cmidrule{2-3} 
\\[-1.8ex] & \multicolumn{2}{c}{ } \\ 
 & WER & Hallucination \\ 
\\[-1.8ex] & (1) & (2)\\ 
\hline \\[-1.8ex] 
 Fluent Aphasia & 0.061*** (0.003) & 1.727*** (0.356) \\
Non-fluent Aphasia & 0.106*** (0.005) & 1.943*** (0.377) \\
Gender (Female) & -0.023*** (0.002) & -0.548*** (0.119) \\
Age & -0.006*** (0.001) & 0.068 (0.057) \\
Age$^2$ & 0.0001*** (0.00001) & -0.0005 (0.0004) \\
< 4-year College & 0.036$^{***}$ (0.004) & 0.079 (0.232) \\
4-year college & 0.049$^{***}$ (0.004) & 0.239 (0.214) \\
Post-grad Degree & 0.036$^{***}$ (0.005) & 0.845$^{***}$ (0.203) \\
Race (Other) & -0.001 (0.009) & 0.300 (0.420) \\
Race (African American) & 0.066$^{***}$ (0.006) & 0.377$^{**}$ (0.190) \\
Mean Background Noise (Audio) & 0.055$^{***}$ (0.008) & 0.336 (0.292) \\
Word Count & 0.0004$^{***}$ (0.00004) & 0.035$^{***}$ (0.010) \\
Total Audio Duration & -0.002$^{***}$ (0.0001) & -0.149$^{***}$ (0.019) \\
Non-vocal Percentage of Audio Duration & 0.242$^{***}$ (0.010) & 2.291$^{***}$ (0.268) \\
ASR (Google Chirp) & 0.018$^{***}$ (0.004) & \\
ASR (Microsoft Azure) & 0.070$^{***}$ (0.004) & \\
ASR (OpenAI Whisper) & 0.029$^{***}$ (0.006) & \\
ASR (AssemblyAI) & 0.048$^{***}$ (0.004) & \\
ASR (Amazon AWS) & 0.046$^{***}$ (0.004) & \\
Constant & 0.162$^{***}$ (0.015) & -8.745$^{***}$ (1.758) \\
 \hline \\[-1.8ex] 
\hline 
\hline \\[-1.8ex] 
\textit{Note:}  & \multicolumn{2}{r}{$^{*}$p$<$0.1; $^{**}$p$<$0.05; $^{***}$p$<$0.01} \\ 
\end{tabular} 
\end{table}

\begin{table}[!h] \centering 
  \caption{Regression analysis of WER and hallucination by the five aphasia types represented in our data sample (Anomic, Conduction, Broca, Wernicke, Global), participant demographic, audio segment features, and ASR with clustered SE on participant (reference levels are the control group, Rev AI, white, less than college education. } 
  \label{tab:regression_hallucination_category} 
\begin{tabular}{@{\extracolsep{5pt}}lcc} 
\\[-1.8ex]\hline 
\hline \\[-1.8ex] 
 & \multicolumn{2}{c}{\textit{Dependent variable:}} \\ 
\cmidrule{2-3} 
\\[-1.8ex] & \multicolumn{2}{c}{ } \\ 
 & WER & Hallucination \\ 
\\[-1.8ex] & (1) & (2)\\ 
\hline \\[-1.8ex] 
Aphasia (Anomic) & 0.046$^{***}$ (0.003) & 1.555$^{***}$ (0.365) \\
Aphasia (Conduction) & 0.086$^{***}$ (0.004) & 1.434$^{***}$ (0.364) \\
Aphasia (Wernicke) & 0.100$^{***}$ (0.007) & 2.582$^{***}$ (0.357) \\
Aphasia (Broca) & 0.116$^{***}$ (0.004) & 2.037$^{***}$ (0.360) \\
Aphasia (Global) & 0.125$^{***}$ (0.016) & 2.983$^{***}$ (0.425) \\
Gender (Female) & -0.015$^{***}$ (0.002) & -0.332$^{***}$ (0.123) \\
Age & -0.006$^{***}$ (0.0004) & 0.114 (0.073) \\
Age 2 & 0.0001$^{***}$ (0.00000) & -0.001 (0.001) \\
< 4-year College & 0.036$^{***}$ (0.004) & 0.254 (0.273) \\
4-year college & 0.049$^{***}$ (0.004) & 0.243 (0.248) \\
Post-grad Degree & 0.034$^{***}$ (0.004) & 0.954$^{***}$ (0.240) \\
Race (Other) & -0.026$^{***}$ (0.007) & -13.043$^{***}$ (0.132) \\
Race (African American) & 0.056$^{***}$ (0.006) & 0.547$^{***}$ (0.197) \\
Mean Background Noise (Audio) & 0.051$^{***}$ (0.006) & 0.324 (0.296) \\
Word Count & 0.0004$^{***}$ (0.00004) & 0.048$^{***}$ (0.008) \\
Total Audio Duration & -0.002$^{***}$ (0.0001) & -0.185$^{***}$ (0.022) \\
Non-vocal Percentage of Audio Duration & 0.238$^{***}$ (0.008) & 2.005$^{***}$ (0.273) \\
ASR (Google Chirp) & 0.017$^{***}$ (0.004) & \\
ASR (Microsoft Azure) & 0.070$^{***}$ (0.004) & \\
ASR (OpenAI Whisper) & 0.024$^{***}$ (0.004) & \\
ASR (AssemblyAI) & 0.047$^{***}$ (0.004) & \\
ASR (Amazon AWS) & 0.044$^{***}$ (0.004) & \\
Constant & 0.150$^{***}$ (0.012) & -10.104$^{***}$ (2.309) \\
 \hline \\[-1.8ex] 
\hline 
\hline \\[-1.8ex] 
\textit{Note:}  & \multicolumn{2}{r}{$^{*}$p$<$0.1; $^{**}$p$<$0.05; $^{***}$p$<$0.01} \\ 
\end{tabular} 
\end{table}

\begin{table}[ht]
\centering
\caption{Linear mixed-effects analysis of WER for unmatched data by aphasia types, participant demographic, audio segment features, and ASR (reference levels are the control group, Rev AI, white, less than college education) }
\label{tab:mixed_effects_model}
\begin{tabular}{p{4.5cm} >{\centering}p{2.2cm} >{\centering}p{2.2cm} >{\centering\arraybackslash}p{2.2cm}}
\hline
Predictors & WER & WER  & WER  \\
\hline
(Intercept) & 0.16 * (0.07) & 0.16 * (0.07) & 0.18 ** (0.06) \\
Aphasia & 0.09 *** (0.01) &  &  \\
Fluent Aphasia &  & 0.07 *** (0.01) &  \\
Nonfluent Aphasia&  & 0.14 *** (0.02) &  \\
Aphasia (Anomic) &  &  & 0.05 *** (0.01) \\
Aphasia (Conduction) &  &  & 0.09 * (0.02) \\
Aphasia (Wernicke) &  &  & 0.12*** (0.02) \\
Aphasia (Broca) &  &  & 0.14 *** (0.01) \\
Aphasia (Global)&  &  & 0.17 ** (0.03) \\
Gender (Female) & -0.04 *** (0.01) & -0.04 *** (0.01) & -0.02 * (0.01) \\
Age & -0.00 (0.00) & -0.00 * (0.00) & -0.00 (0.00) \\
Age$^2$ & 0.00 * (0.00) & 0.00 * (0.00) & 0.00 * (0.00) \\
2-year College or dropped out of 4-year College & 0.01 (0.02) & 0.01 (0.02) & 0.02 (0.02) \\
4-year College & 0.03 (0.02) & 0.03 * (0.02) & 0.04 ** (0.01) \\
Post-grad Degree & 0.01 (0.02) & 0.01 (0.02) & 0.01 (0.01) \\
Race (Other) & 0.00 (0.03) & -0.00 (0.03) & -0.03 (0.03) \\
Race (African American) & 0.01 (0.02) & 0.03 (0.02) & 0.02 (0.02) \\
Audio Mean Background Noise & 0.01 (0.01) & -0.00 (0.02) & 0.00 (0.01) \\
Word Count & 0.00 * (0.00) & 0.00 ** (0.00) & 0.00 ** (0.00) \\
Total Audio Duration & -0.00 *** (0.00) & -0.00 *** (0.00) & -0.00 *** (0.00) \\
Nonvocal Percentage Share (\%) & 0.20 *** (0.01) & 0.21 *** (0.01) & 0.21 *** (0.01) \\
ASR (Google Chirp) & 0.02 *** (0.00) & 0.02 *** (0.00) & 0.02 *** (0.00) \\
ASR (Microsoft Azure) & 0.07 *** (0.00) & 0.07 *** (0.00) & 0.07 *** (0.00) \\
ASR (OpenAI Whisper) & 0.03 *** (0.00) & 0.03 *** (0.00) & 0.02 *** (0.00) \\
ASR (AssemblyAI) & 0.05 *** (0.00) & 0.05 *** (0.00) & 0.05 *** (0.00) \\
ASR (Amazon AWS) & 0.05 *** (0.00) & 0.05 *** (0.00) & 0.04 *** (0.00) \\

\hline
\multicolumn{4}{l}{\textbf{Random Effects}} \\
$\sigma^2$ & 0.03 & 0.03 & 0.02 \\
$\tau_{00}$ segment\_name & 0.02 & 0.02 & 0.02 \\
$\tau_{00}$ Participant\_ID & 0.01 & 0.01 & 0.01 \\
ICC & 0.53 & 0.51 & 0.65 \\
N Participant\_ID / segment\_name & 537 / 6185 & 505 / 5681 & 466 / 5337 \\
Observations & 37110 & 34086 & 32022 \\
Marginal R$^2$ / Conditional R$^2$ & 0.114 / 0.581 & 0.130 / 0.573 & 0.175 / 0.708 \\
\hline
\multicolumn{4}{l}{\small{* p<0.05, ** p<0.01, *** p<0.001}}
\end{tabular}
\end{table}

\newpage

\subsection{Evaluation Metrics and Hallucination Detection}

This section supports the discussion in \autoref{sec:pitfall-relying_on_wer} on the limitations of WER and the importance of detecting hallucinations.

\subsubsection{Suite of Evaluation Metrics}
\autoref{tab:asr_metrics} presents performance across a comprehensive suite of automated metrics beyond WER, including BLEU, ROUGE, METEOR, CER, WIL, and RIL.

\begin{landscape}
\begin{table}[ht]
\centering
\caption{Suite of average automated ASR metrics (higher is better for BLEU, ROUGE-1, ROUGE-2, ROUGE-L, and METEOR; lower is better for CER, Insertion, WIL, and RIL). For each metric, the best performing service is marked in bold; all services perform better on the control group than the aphasia group on all metrics. These metrics were used to determine the subset of 1,198 audio files on which manual review was performed to check for hallucinations.}
\begin{tabular}{lccccccccc}
\toprule
\textbf{ASR Service} & \textbf{BLEU} & \textbf{CER} & \textbf{Insertion} & \textbf{ROUGE-1} & \textbf{ROUGE-2} & \textbf{ROUGE-L} & \textbf{WIL} & \textbf{RIL} & \textbf{METEOR} \\
\midrule
Rev AI                &               &              &                    &                  &                  &                  &              &              &              \\
\quad Aphasia        & 0.70          & 0.15         & 0.05               & 0.87             & 0.79             & 0.87             & 0.22         & 0.18         & 0.85         \\
\quad Control        & \textbf{0.85}          & \textbf{0.07}        & 0.02               & \textbf{0.94}             & \textbf{0.89}             & \textbf{0.94}             & \textbf{0.12}         & \textbf{0.09}         & \textbf{0.93}         \\
Google Chirp         &               &              &                    &                  &                  &                  &              &              &              \\
\quad Aphasia        & 0.67          & 0.16         & 0.03               & 0.85             & 0.76             & 0.85             & 0.24         & 0.18         & 0.82         \\
\quad Control        & 0.82          & 0.08         & 0.02               & 0.93             & 0.87             & 0.92             & 0.14         & 0.10         & 0.91         \\
OpenAI Whisper       &               &              &                    &                  &                  &                  &              &              &              \\
\quad Aphasia        & 0.67          & 0.18         & 0.05               & 0.85             & 0.76             & 0.85             & 0.25         & 0.34         & 0.81         \\
\quad Control        & \textbf{0.85}         & 0.08         & 0.02               & 0.94             & 0.89             & 0.93             & \textbf{0.12}         & 0.10         & 0.92         \\
Amazon AWS           &               &              &                    &                  &                  &                  &              &              &              \\
\quad Aphasia        & 0.63          & 0.19         & 0.05               & 0.83             & 0.72             & 0.83             & 0.28         & 0.24         & 0.80         \\
\quad Control        & 0.81          & 0.09         & 0.02               & 0.92             & 0.86             & 0.92             & 0.15         & 0.11         & 0.90         \\
Microsoft Azure      &               &              &                    &                  &                  &                  &              &              &              \\
\quad Aphasia        & 0.61          & 0.21         & 0.05               & 0.81             & 0.70             & 0.81             & 0.31         & 0.24         & 0.77         \\
\quad Control        & 0.78          & 0.11         & 0.02               & 0.90             & 0.84             & \textbf{0.89}             & 0.18         & 0.12         & 0.87         \\
AssemblyAI           &               &              &                    &                  &                  &                  &              &              &              \\
\quad Aphasia        & 0.63          & 0.20         & 0.03               & 0.83             & 0.73             & 0.83             & 0.28         & 0.20         & 0.77         \\
\quad Control        & 0.81          & 0.10         & \textbf{0.01}               & 0.92             & 0.87             & 0.91             & 0.15         & 0.10         & 0.89         \\
\bottomrule
\end{tabular}
\label{tab:asr_metrics}
\end{table}
\end{landscape}

\subsubsection{Distinction Between Hallucination and Mistranscription}\label{sec:hallucination-pipeline}

To identify hallucinations within Whisper-generated ASR transcriptions, we use a combined computational and manual review approach. First, we computationally assess whether the ASR transcription is identical to the ground truth, or if only deletions are present without any substitutions or insertions. If this is the case, the transcription is classified as mistranscription. Next, we use a set of NLP-based metrics to computationally select a focused subset of audio files for manual review (described in the following paragraph). Manual checks are for one or more clear indicators of hallucination traits, including perpetuation of violence, inaccurate association, false authority, random semantically irrelevant words/phrases, repetitions not present in the audio file, made-up words or phrases that continue after the end of the sentence or clause, and language switching (following taxonomies identified in prior work~\cite{koenecke2024careless}). If no such traits are present, the output is considered a mistranscription. However, if hallucination traits are detected, the audio file is listened to---to manually to confirm no potential influence of phonetic features, contextual coarticulation, and measurable acoustic parameters (like formants, amplitude, and duration) on the hallucinated text. If such features exist, the error is considered  a mistranscription; else, it is classified as a hallucination. 

We employed a set of NLP-based metrics to computationally select a focused subset of audio files for manual review (Bilingual Evaluation Understudy (BLEU), Recall-Oriented Understudy for Gisting Evaluation (ROUGE), METEOR, WIL, RIL, CER, and Insertion Rate); average rates are calculated for all audio snippets used for main text analyses and are reported in Table 3. To isolate the most extreme values where hallucinations are likely to occur, we use percentile-based thresholds---selecting values above the 90th percentile for metrics where lower is better (like WER) and below the 10th percentile for metrics where higher is better (like most NLP-related scores)---narrowing down to 1,198 high-priority candidate files. Among these, we manually identified 56 files containing Whisper hallucinations. Of these files, 35 contained hallucinations of non-English languages, 16 contained hallucinated content including YouTube captioning or website text that bears no relation to the original audio, 2 contained attempts to complete or extend utterances in semantically plausible ways despite no such continuations being present in the audio, and 4 contained repeated phrases. Spot checks conducted throughout the cleaning process confirmed that, among the files manually examined---focusing on transcriptions with high-risk indicators similar to those found in Whisper data---only Whisper transcriptions exhibited these hallucinations.

\subsubsection{Hallucination Experiment Results}\label{sec:hallucination-results}

\begin{landscape}
\begin{figure}[htbp]
    \centering
    \includegraphics[width=\textwidth]{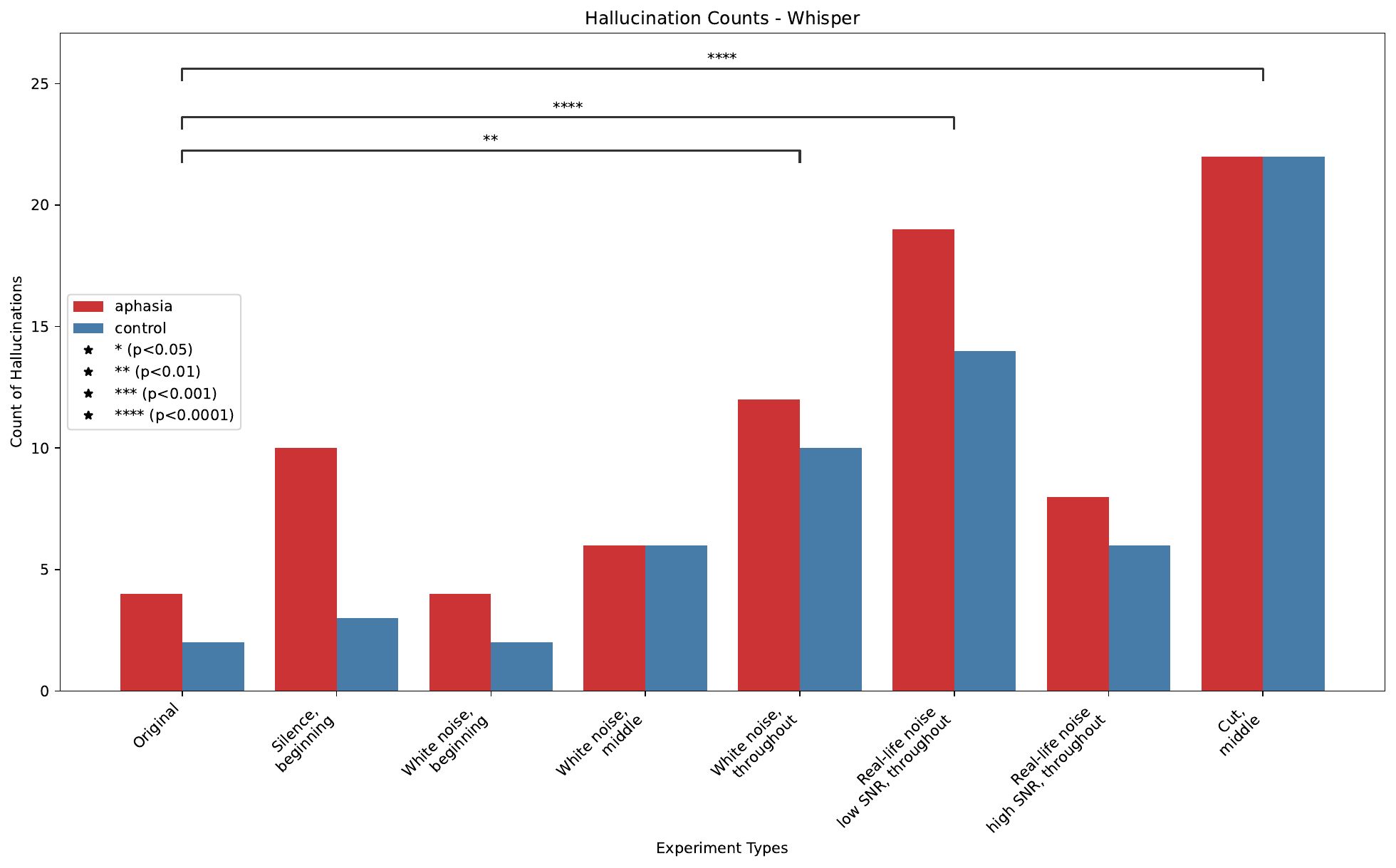}
    \caption{Hallucinations found for each experiment arm in aphasia and control groups. McNemar's test was conducted to compare the results between the original audio file and the manipulated counterpart. Experiment arms that have statistically significant differences from the original audio are noted.}
    \label{fig:hallucination-count}
\end{figure}
\end{landscape}

We additionally conducted an experiment to observe whether certain inserted audio file characteristics would be more likely to trigger a Whisper hallucination. All audio files were run through the OpenAI Whisper API once more during the last week of February, 2024. We created seven different experimental treatments that directly manipulated audio file characteristics: (a) inserting 10 seconds of silence at the beginning of the audio file, (b) inserting 1 second of white noise at the beginning, (c) inserting 1 second of white noise mid-sentence, (d) inserting white noise throughout, (e) inserting real-life noise throughout with two different levels of Signal-to-Noise Ratio (SNR), and (f) cutting the audio file before the utterance was completed. To add silence, we used the AudioSegment package from Python, creating a silent audio segment of 10,000 ms and appending the segment to the beginning of an audio file. To add white noise, we used the Numpy package from Python to generate white noise with a Gaussian distribution, which has equal power across all frequencies within the defined bandwidth. We created 1,000 ms of white noise at the level of 5 dB SNR, which means the average power of the signal, the original audio file in this case, is 5 dB greater than the generated white noise, and appended the noise to either the beginning or end of an audio file. To add real-life noise, we used a real-life noise file obtained online (\url{https://freesound.org/people/eguobyte/sounds/360703/}) and modified the file to be 5 dB SNR. After initial experiments with this setting, we noticed that the WER was extremely high for files with this treatment and the audio files were, in the extreme cases, inaudible to human ears as well. We hence added another experiment with the same setting, except for having the noise at a higher 10 dB SNR to ensure speech intelligibility. For treatments (c) and (f), an arbitrary insertion point that is positioned mid-sentence, but not mid-word, was selected after manually listening and verifying each audio file, making sure to mask at least three words from the original audio file. The same point was used for both treatments. 241 audio files were selected for our experiment using stratified sampling based on demographic factors including gender, age, education level, and word count, with race removed as a stratification factor due to underrepresentation of certain subgroups. A total of 1,928 audio files were newly transcribed using OpenAI Whisper, from which we identified 150 hallucinations. This is an extremely high proportion (7.8\%) relative to the previously observed 0.9\% hallucination rate from the original audio files without acoustic manipulation. As shown in \autoref{fig:hallucination-count}, three conditions---inserting white noise and real-life noise throughout the audio file at a low SNR, and cutting the audio file early---impacted the occurrence of hallucinations at a statistically significant level. Specifically, prematurely cutting audio files resulted in hallucinations of one or two words that seemed to infer words after the cut point. While none of the experiment arms resulted in statistically significant differences in hallucination generation between aphasia and control groups, the count of hallucination occurrences was higher for the aphasia group.

\end{document}

%% file: tables/demographic_table.tex
\begin{table}[h!]
\centering
\footnotesize
\caption{Participant Demographics by Group}
\resizebox{0.6\columnwidth}{!}{%
\begin{tabular}{llrr}
\toprule
Demographic & Category & Control & Aphasia \\
\midrule

\multicolumn{4}{l}{\textbf{Gender}} \\
 & Female & 124 & 134 \\
 & Male   & 92  & 187 \\
\midrule

\multicolumn{4}{l}{\textbf{Race}} \\
 & White (WH) & 207 & 277 \\
 & African American (AA) & 4 & 34 \\
 & Hispanic/Latino (HL) & 5 & 6 \\
 & Asian (AS) & -- & 3 \\
 & Other (OTH) & -- & 1 \\
\midrule

\multicolumn{4}{l}{\textbf{Primary Language}} \\
 & English & 216 & 321 \\
\midrule

\multicolumn{4}{l}{\textbf{Employment Status}} \\
 & Retired (Retired or Not working) & 76 & 276 \\
 & Retired, Working (R/W) & 36 & -- \\
 & Working (W) & 19 & 21 \\
 & Not Available & 85 & 24 \\

\midrule
\multicolumn{4}{l}{\textbf{Age Group}} \\
&18-20 & 1 & --\\
&21-30 & 10 & 3\\
&31-40 & 16 & 9\\
&41-50 & 21 & 44\\
&51-60 & 36 & 81\\
&61-70 & 33 & 96\\
&71-80 & 69 & 72\\
&Over 80 & 30 & 16\\
\midrule

\multicolumn{4}{l}{\textbf{Adequate Hearing}} \\
 & Yes & 215 & 318 \\
 & No  & 1   & 2 \\
 & Not Available & -- & 1 \\
\midrule

\multicolumn{4}{l}{\textbf{Adequate Vision}} \\
 & Yes & 216 & 318 \\
 & No  & --  & 3 \\
\midrule

\multicolumn{4}{l}{\textbf{Aphasia Category (Clinical Impression)}} \\
 & Fluent (FLU) & -- & 157 \\
 & Non-Fluent (NFL) & -- & 129 \\
 & Latent & -- & 5 \\
 & NCL & -- & 7 \\
 & Not Available & -- & 23 \\
\midrule
\multicolumn{4}{l}{\textbf{Aphasia Type (Boston Classification, Clinical Impression)}} \\
 & Anomic (ANO) & -- & 98 \\
 & Broca (BRO) & -- & 89 \\
 & Conduction (CON) & -- & 32 \\
 & Global (GLO) & -- & 12 \\
 & Mixed Transcortical (MTC) & -- & 1 \\
 & Non-Classifiable (NCL) & -- & 13 \\
 & Optic (OPT) & -- & 1 \\
 & Other (OTH) & -- & 1 \\
 & Transcortical Motor (TCM) & -- & 7 \\
 & Transcortical Sensory (TCS) & -- & 3 \\
 & Wernicke (WER) & -- & 19 \\
 & Not Available & -- & 45 \\
\bottomrule
\end{tabular}}
\label{tab:overall_participant_demographics}
\end{table}

%% file: tables/distributional_balance_table.tex
\begin{table}[!h]
\caption{Demographic distribution counts of audio files and participants before and after propensity matching. Here, ``Other'' Race/Ethnicity combines Hispanic/Latino, Asian, and Other race categories due to small counts therein.}
\centering
\begin{tabular}[t]{p{4cm}|p{1.5cm}p{1.5cm}|p{1.5cm}p{1.5cm}}
\toprule
Demographics & \multicolumn{2}{c|}{Audio Files} & \multicolumn{2}{c}{Participants} \\
\midrule
 & Pre-matching & Post-matching & Pre-matching & Post-matching \\
\midrule
\multicolumn{5}{l}{\textbf{Race/Ethnicity}} \\
\midrule
\hspace{1em}African American & 486 & 64 & 38 & 18 \\
\hspace{1em}Other & 128 & 59 & 15 & 13 \\
\hspace{1em}White & 5571 & 2325 & 484 & 426 \\
\midrule
\multicolumn{5}{l}{\textbf{Gender}} \\
\midrule
\hspace{1em}Male & 3043 & 1055 & 279 & 223 \\
\hspace{1em}Female & 3142 & 1393 & 258 & 234 \\
\midrule
\multicolumn{5}{l}{\textbf{Education Levels}} \\
\midrule
\hspace{1em}$\leq$ High School Degree & 1259 & 474 & 102 & 83 \\
\hspace{1em}< 4-year College & 1261 & 490 & 112 & 93 \\
\hspace{1em}4-year college & 1823 & 761 & 163 & 142 \\
\hspace{1em}Post-grad Degree & 1842 & 723 & 160 & 139 \\
\midrule
\multicolumn{5}{l}{\textbf{Age}} \\
\midrule
\hspace{1em}18-20 & 46 & 37 & 1 & 1 \\
\hspace{1em}21-29 & 254 & 155 & 13 & 12 \\
\hspace{1em}31-40 & 181 & 93 & 25 & 23 \\
\hspace{1em}41-50 & 786 & 245 & 65 & 55 \\
\hspace{1em}51-60 & 1298 & 489 & 117 & 91 \\
\hspace{1em}61-70 & 1638 & 576 & 129 & 111 \\
\hspace{1em}71-80 & 1591 & 654 & 141 & 121 \\
\hspace{1em}Over 80 & 391 & 199 & 46 & 43 \\
\bottomrule
\end{tabular}
\label{tab:pre-post-matching-count}
\end{table}

%% file: tables/love_plot.tex
\begin{figure}[!th]
    \centering    \includegraphics[width=0.5\linewidth]{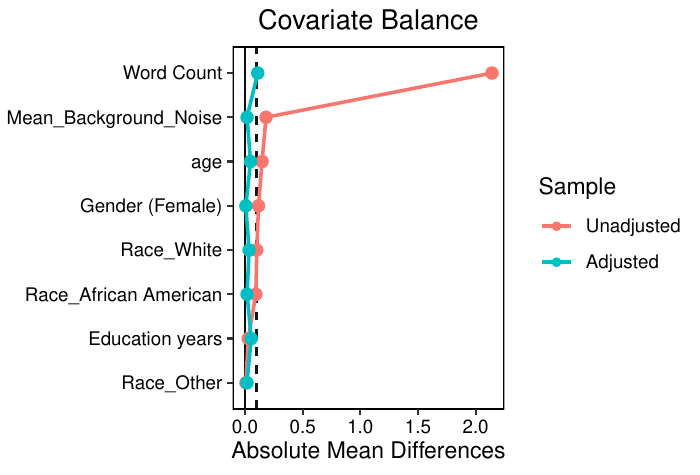}
    \caption{Love plot for standardized mean differences for covariates before and after propensity matching: as shown in the plot, absolute mean differences for covariates are below 0.1 (dotted line) after matching.}
    \label{fig:loveplot}
\end{figure}

%% file: tables/kruskal_wallis_result.tex
\begin{table}[h!]
\centering
\caption{Kruskal–Wallis test results assessing differences among transcript versions across ASR services. }
\resizebox{0.7\columnwidth}{!}{%
\begin{tabular}{lrrrr}
\toprule
\textbf{ASR} 
& \multicolumn{2}{c}{\textbf{Aphasia}} 
& \multicolumn{2}{c}{\textbf{Control}} \\
\cmidrule(lr){2-3} \cmidrule(lr){4-5}
 & $\chi^2$ & $p$-value & $\chi^2$ & $p$-value \\
\midrule
Amazon AWS        & 5242.28 & $<.001$ &  2532.54 & $<.001$ \\
Microsoft Azure   & 5650.74 & $<.001$ &  2552.37 & $<.001$ \\
AssemblyAI        & 5471.39 & $<.001$ &  2787.92 & $<.001$ \\
Google Chirp      & 1432.29 & $<.001$ & 848.75  &$<.001$    \\
Rev AI            & 8259.88 & $<.001$ & 3464.16 & $<.001$ \\
OpenAI Whisper    & 5329.59 & $<.001$ &  2921.64 & $<.001$ \\
\bottomrule
\end{tabular}}
\label{tab:kw_test}
\end{table}

%% file: tables/mapsswe_test_result_aphasia.tex
\begin{table}
\caption{MAPSSWE test results comparing WER between ASR services across different standardization levels for the aphasia group. Each cell shows the mean difference in WER and the corresponding W statistic. Rows shaded in light gray indicate that the ranking of services changed with the standardization level. Significance: $*: p<0.05$.}
\centering
\footnotesize
\begin{tabular}{p{4cm}p{0.8cm}p{0.8cm}p{0.8cm}p{0.8cm}p{0.5cm}p{0.8cm}p{0.8cm}p{0.8cm}p{0.8cm}p{0.8cm}}
\toprule
Comparison & 
\multicolumn{2}{c}{Original} & 
\multicolumn{2}{c}{Remove fillers} & 
\multicolumn{2}{c}{\makecell{Remove fillers \\and fragments}} & 
\multicolumn{2}{c}{\makecell{Remove fillers, \\fragments,\\ and repeated words}} & 
\multicolumn{2}{c}{\makecell{Remove fillers, \\fragments, \\repeated words, \\and repeated phrases}} \\
\cmidrule(lr){2-3} \cmidrule(lr){4-5} \cmidrule(lr){6-7} \cmidrule(lr){8-9} \cmidrule(lr){10-11}
 & Mean Diff & W  & Mean Diff & W  & Mean Diff & W & Mean Diff & W  & Mean Diff & W  \\
\midrule
Google Chirp vs Microsoft Azure & -0.32&-87.33* & -0.05&-20.96* & -0.06&-20.41* & -0.06&-18.65* & -0.06&-18.1*\\
Google Chirp vs OpenAI Whisper & -0.25&-62.88* & -0.04&-9.46* & -0.04&-9.51* & -0.03&-4.84* & -0.02&-3.39*\\
\rowcolor{gray!30}
Google Chirp vs Rev AI & -0.46&-90.77* & 0.02&10.8* & 0.02&8.71* & 0.02&7.6* & 0.02&7.4*\\

Google Chirp vs AssemblyAI & -0.29&-84.46* & -0.07&-27.49* & -0.07&-24.83* & -0.05&-16.82* & -0.04&-12.92*\\

Google Chirp vs Amazon AWS & -0.26&-75.93* & -0.03&-11.47* & -0.03&-12.63* & -0.04&-12.25* & -0.04&-12.12*\\

Microsoft Azure vs OpenAI Whisper & 0.07&21.6* & 0.01&3.39* & 0.02&3.57* & 0.03&3.97* & 0.04&6.01*\\
\rowcolor{gray!30}
Microsoft Azure vs Rev AI & -0.14&-40.24* & 0.07&31.43* & 0.08&28.93* & 0.08&25.22* & 0.08&24.49*\\
\rowcolor{gray!30}
Microsoft Azure vs AssemblyAI & 0.03&13.71* & -0.02&-7.68* & -0.01&-4.18* & 0.01&2.75* & 0.02&6.42*\\

Microsoft Azure vs Amazon AWS & 0.06&26.93* & 0.02&10.28* & 0.02&9.08* & 0.02&7.65* & 0.02&7.11*\\
\rowcolor{gray!30}
OpenAI Whisper vs Rev AI & -0.21&-53.17* & 0.06&15.63* & 0.06&14.71* & 0.05&7.79* & 0.04&6.52*\\

OpenAI Whisper vs AssemblyAI & -0.04&-12.38* & -0.04&-8.6* & -0.03&-6.29* & -0.02&-2.66* & -0.02&-2.68*\\
\rowcolor{gray!30}
OpenAI Whisper vs Amazon AWS & -0.01&-3.42* & 0.01&2.49* & 0.01& 1.85  & 0 &-0.58  & -0.02& -2.54*\\
\rowcolor{gray!30}
Rev AI vs AssemblyAI & 0.18&51.2* & -0.1&-37.96* & -0.09&-32.52* & -0.07&-23.49* & -0.06&-19.6*\\
\rowcolor{gray!30}
Rev AI vs Amazon AWS & 0.2&63.46* & -0.05&-22.48* & -0.05&-21.59* & -0.05&-19.78* & -0.05&-19.11*\\

AssemblyAI vs Amazon AWS & 0.03&12.56* & 0.05&16.26* & 0.04&11.95* & 0.01&4.27* & 0&0.27\\
\bottomrule
\end{tabular}
\label{tab:mapsswe_aphasia}
\end{table}